\documentclass[%aps,prl,
aip, jpc, reprint,noshowkeys,superscriptaddress]{revtex4-1}
\usepackage{graphicx,dcolumn,bm,xcolor,microtype,multirow,amscd,amsmath,amssymb,amsfonts,physics,wrapfig,bbold,siunitx,xspace}
\usepackage[version=4]{mhchem}

\usepackage[utf8]{inputenc}
\usepackage[T1]{fontenc}

\usepackage{hyperref}
\hypersetup{
    colorlinks,
    linkcolor={red!50!black},
    citecolor={red!70!black},
    urlcolor={red!80!black}
}

\usepackage{listings}
\definecolor{codegreen}{rgb}{0.58,0.4,0.2}
\definecolor{codegray}{rgb}{0.5,0.5,0.5}
\definecolor{codepurple}{rgb}{0.25,0.35,0.55}
\definecolor{codeblue}{rgb}{0.30,0.60,0.8}
\definecolor{backcolour}{rgb}{0.98,0.98,0.98}
\definecolor{mygray}{rgb}{0.5,0.5,0.5}

\definecolor{sqred}{rgb}{0.85,0.1,0.1}
\definecolor{sqgreen}{rgb}{0.25,0.65,0.15}
\definecolor{sqorange}{rgb}{0.90,0.50,0.15}
\definecolor{sqblue}{rgb}{0.10,0.3,0.60}

\lstdefinestyle{mystyle}{
    backgroundcolor=\color{backcolour},
    commentstyle=\color{codegreen},
    keywordstyle=\color{codeblue},
    numberstyle=\tiny\color{codegray},
    stringstyle=\color{codepurple},
    basicstyle=\ttfamily\footnotesize,
    breakatwhitespace=false,
    breaklines=true,
    captionpos=b,
    keepspaces=true,
    numbers=left,
    numbersep=5pt,
    numberstyle=\ttfamily\tiny\color{mygray},
    showspaces=false,
    showstringspaces=false,
    showtabs=false,
    tabsize=2
  }

  \newcolumntype{d}{D{.}{.}{-1}}

\usepackage{graphicx}% Include figure files
\usepackage{dcolumn}% Align table columns on decimal point
\usepackage{bm}% bold math
\usepackage{scalerel,stackengine}
\usepackage{multirow}
\usepackage{xcolor}
\usepackage{caption}
\usepackage{subcaption}
\usepackage{scalerel,stackengine}
\stackMath
\newcommand\reallywidehat[1]{%
\savestack{\tmpbox}{\stretchto{%
  \scaleto{%
    \scalerel*[\widthof{\ensuremath{#1}}]{\kern.1pt\mathchar"0362\kern.1pt}%
    {\rule{0ex}{2\textheight}}%WIDTH-LIMITED CIRCUMFLEX
  }{\textheight}% 
}{2.4ex}}%
\stackon[-6.5pt]{#1}{\tmpbox}%
}

\def \lam{\lambda}
\def \hf{\text{HF}}
\def \ks{\text{KS}}

\def \ron{n}

\newcommand{\ud}{\mathrm{d}}
\newcommand{\rv}{\mathbf{r}}
\def \br{\mathbf{r}}
\def \bx{\mathbf{x}}
\DeclareMathOperator*{\argmax}{arg\,max}
\DeclareMathOperator*{\argmin}{arg\,min}

\newcommand{\changes}[1]{\textcolor{black}{#1}}

\begin{document}

%\preprint{APS/123-QED}

\title{M\o ller-Plesset  and density-fixed adiabatic connections for a model diatomic system at different correlation regimes}

\author{Sara Giarrusso}
 \email{sgiarrusso@ucmerced.edu}
% \altaffiliation[Also at ]{Department of Chemistry and Biochemistry, University of California Merced, 5200 North Lake Rd. Merced, CA 95343, USA.}%Lines break automatically or can be forced with \\
\author{Aurora Pribram-Jones}%
 \email{apj@ucmerced.edu}
\affiliation{%
Department of Chemistry and Biochemistry, University of California Merced, 5200 North Lake Rd. Merced, CA 95343, USA
}%

\date{\today}% It is always \today, today,

\begin{abstract}
%Adiabatic Connection Interpolation methods in electronic structures have proven useful for the calculation of binding and interaction energies, but there is ongoing effort to improve them further.
In recent years, Adiabatic Connection Interpolations developed within Density Functional Theory (DFT) have been found to provide good performances in the calculation of interaction energies when used with Hartree-Fock (HF) ingredients.
The physical and mathematical reasons for such unanticipated performance have been clarified, to some extent, by studying the strong-interaction limit of the M\o ller-Plesset (MP) adiabatic connection.
%Nonetheless, the MP adiabatic connection appears to be more flexible that the DFT one and there is not as many example of its accurate calculation in simple systems or as many exact results as those concerning the DFT adibatic connection that can guide the development of interpolation formulas specifically designed for it.
In this work, we calculate both the MP and the DFT adiabatic connection (AC) integrand for the asymmetric Hubbard dimer, which allows for a systematic investigation at different correlation regimes by varying two simple parameters in the Hamiltonian: the external potential, $\Delta v$, and the interaction strength, $U$. Noticeably, we find that, while the DFT AC integrand appears to be convex in the full parameter space, the MP integrand may change curvature twice. 
Furthermore, we discuss different aspects of the second-order expansion of the correlation energy in each adiabatic connection %, such as their link to the accuracy predictor proposed in reference~\cite{VucFabGorBur-JCTC-20}.Next, 
and we demonstrate \changes{why} the derivative of the $\lam$-dependent density in the MP adiabatic connection at $\lam=0$ (i.e., at the HF density) is zero \changes{in the model}. 
Concerning the strong-interaction limit of both adiabatic connections, we show that while, for a given density, the asymptotic value of the MP adiabatic connection, $W_\infty^\hf$, is lower (or equal) than its DFT analogue, $W_\infty^\ks$,  this is not always the case for a given external potential.
%While the model is extremely simple, it is possible that some of these newly-observed features are shared with actual molecular systems.

%while the model is extremely simplified, it allows for an interesting discussion of correlation and correlation regimes, which is exhaustive in the model.
\end{abstract}

\maketitle

%\tableofcontents
\section{Introduction and Theoretical Background}
Adiabatic connection methods rely on the idea of gradually switching from a formally non-interacting Hamiltonian, which is comparatively simple, to a ``fully interacting'' one, which is more complicated and describes the actual electronic system of interest. This is done by multiplying the interaction operator by a coupling- or interaction-strength parameter.

 Nowadays, there are several different flavors of adiabatic connections adopted in wavefunction-based methods;\cite{Per-PRL-18, Per-IJQC-18, PasPer-JCTC-18,Per-JCP-18, PasPer-JPCL-18, PasHapVeiPer-JPCL-19, DrwPasPer-JCP-21,BerMatHapPerVei-JCTC-21,DrwBerHapModSokVeiPer-JPCL-22, MatHapVeiPer-JCP-23}however, the formalism was first developed in the context of Density Functional Theory (DFT)~\cite{HarJon-JPF-74, GunLun-PRB-76, LanPer-SSC-75} and it has been a quite powerful tool to construct models for the exchange-correlation (XC) energy in Kohn-Sham DFT~\cite{KS65} ever since. %A key feature of the DFT adiabatic connection is that the density is constrained along the different coupling strengths.  Even more importantly, the density-fixed adiabatic connection is less amenable to a perturbative expansion at small $\lam$, because the density-constraint introduces a $\lam$-dependence in the external potential operator that enters order by order as a ``perturbation'' together with the two-body operator~\cite{GorLev-PRB-93, GorLev-PRA-94}. 
%Thus, typically in this context, only the zeroth- or first-order terms of the small $\lam$ expansion are used. 
Indeed, it has provided the rationale for density functional approximations (DFAs) such as hybrid,\cite{Bec-JCP-93, PerBurErn-JCP-96} double-hybrid,\cite{ShaTouSav-JCP-11}  and functionals from the random phase approximation,\cite{Fur-PRB-01} encoding the exact behaviour of the adiabatic connection at small coupling, where the interaction can be treated as a perturbation.

In addition to these types of functionals however, this formalism has also inspired the construction of DFAs that interpolate between two different limits of the adiabatic connection curve%(known or estimated) and model this latter directly
, performing what is effectively an (approximate) all-order resummation of the perturbation series.
Initially, DFAs of this kind were built as an interpolation between the zero- and the full-interaction limits,\cite{Ern-CPL-96} but, shortly afterwards, more balanced interpolations were constructed by extending the range of the coupling strength \textit{beyond} the physical value, bringing it to infinity\cite{SeiPerLev-PRA-99, SeiPerKur-PRL-00, LiuBur-PRA-09, GorVigSei-JCTC-09, Con-PRB-19} and thus combining the information coming from equally extreme limits (where equally extreme is in the sense that the coupling-strength parameter $\lambda$ in front of the interaction operator in the two limits behaves as $\lambda \to \infty$ or $\alpha \to \infty$ with $\alpha=\frac{1}{\lambda}$).  % (if one chooses to see it under this light).
This class of DFAs, collectively referred to as Adiabatic Connection (Interaction) Interpolations (ACIIs or ACIs) or Adiabatic Connection Methods (ACMs), has recently drawn much attention. One important reason for such renewed interest is that their lack of size-consistency can be corrected very easily at no extra computational cost, as shown in Ref.~\onlinecite{VucGorDelFab-JPCL-18}. Another fundamental reason is that, although having been originally devised in a DFT framework, ACMs have been shown to provide satisfactory performances for binding and interaction energies (in non-covalent complexes), when used with Hartree-Fock (HF) ingredients.\cite{FabGorSeiDel-JCTC-16, VucGorDelFab-JPCL-18,  VucFabGorBur-JCTC-20, DaaFabDelGorVuc-JPCL-21}

 Their use in this framework has numerous practical advantages compared to their use in KS-DFT and some theoretical downsides. The downside compared to KS-DFT is that ACMs on HF ingredients are a simple energetic correction to the HF approximation: they  cannot be used to obtain the interacting density via a self-consistent-field (SCF) scheme. On the contrary, ACMs within KS-DFT can in principle yield an approximate interacting density via an SCF calculation, but with the practical disadvantage that their implementation 
is quite involved and expensive, due to the presence of functional derivatives \changes{of energy terms that depend only implicitly on the density}.\cite{FabSmiGiaDaaDelGraGor-JCTC-19, SmiDelGorFab-JCTC-22} %and loses the computational advantage of KS-DFT over wavefunction-based methods. 
Indeed, ACMs within DFT have been mostly used on approximate KS orbitals, obtained from a preceding SCF calculation, but this strategy introduces an extra layer of approximation, falls back into the known problem of having to ``cherry-pick'' the best functional for the calculation at hand, and seems to be overall not quite accurate.\cite{FabGorSeiDel-JCTC-16}

By contrast, when using ACMs within the MP adiabatic connection, the orbitals required from the theory are simply the HF orbitals, which are fixed once for all in the initial HF calculation. %This huge simplification over the KS-DFT framework naturally comes at some cost: while in principle within the SCF KS-DFT framework (these ACMs offer a strategy for the calculation of the interacting density as well as the energy), the ACMs on HF ingredients are meant to be an intrinsically post-SCF and purely energetic correction to the HF approximation.
This use of interpolation formulas with the HF density and orbitals 
has been theoretically supported by studies on the strong-interaction limit of the associated adiabatic connection~\cite{SeiGiaVucFabGor-JCP-18, DaaKooGroSeiGor-JCTC-22} (see also Ref.~\onlinecite{VucGerDaaBahFriGor-WIRE-22} for a review). 

Although these theoretical advancements pave the way for the use of HF ingredients in a density-functional spirit, many things still need to be better understood. For instance, the ACMs were constructed as convex interpolants, since the DFT adiabatic connection is reckoned to be (piecewise) convex. However, the MP adiabatic connection is known to be concave in the small-interaction region for some simple atomic and molecular systems,\cite{Per-JCP-18, VucFabGorBur-JCTC-20,DaaGroVucMusKooSeiGieGor-JCP-20} therefore some variations of ACMs had to be developed to accommodate this feature.\cite{DaaFabDelGorVuc-JPCL-21}
To further the development of ACMs in both frameworks, we present in this work a careful comparison between the MP and the DFT adiabatic connection for a model diatomic system at different correlation regimes: the asymmetric Hubbard dimer. 
This model has been useful in the context of Site-Occupation Function Theory (SOFT), the analogue of DFT for lattice systems, %and is particularly suited to study the dependence of relevant quantities in electronic structure theories on the density 
because the density (or ``site-occupation difference'' in the model) can be varied easily by varying two simple parameters in the model Hamiltonian.
%This model has been useful in DFT, or more precisely in the context of Site-Occupation Function Theory (SOFT), which is the analogous of DFT for lattice systems, because it is easy to express relevant quantities in electronic structure theories (energies, potentials, molecular properties) as functions of the density (embodied by the site-occupation difference in the model)

In the following, we review the theory of the MP and DFT adiabatic connections (section~\ref{sec:ACs}) as well as the model system (section~\ref{sec:HD}). Section~\ref{sec:ACforHD} translates the two adiabatic connections in the language of the Hubbard dimer, while
section~\ref{sec:Results} illustrates the results of the calculations: the shapes of the two curves at different points in the parameter space (section~\ref{sec:lACI}), the performance of a proposed indicator~\cite{VucFabGorBur-JCTC-20} as a predictor of the accuracy of the correlation energy expanded up to second-order (section~\ref{sec:accpred}) and the density as a function of the coupling parameter in the MP adiabatic connection (section~\ref{sec:lSOD}).
Section~\ref{sec:SIL} focuses entirely on the strong-interaction limit of the two adibatic connections, while section~\ref{sec:Conclu} gives some conclusive remarks. %and outlook on future work.

% This model is popular in electronic structure as a playground to test methods or study their features due to its simplicity combined with a comparably remarkable predictive power in terms of properties of actual electronic structures.

%this strategy gave important results and more recently --> full circle --> HF

%Hartree-Fock is amazing...

%Maybe put here the plot of the asymptotic expansions? For motivating/conextualizing purposes

\subsection{M\o ller-Plesset and density-fixed adiabatic connections}\label{sec:ACs}%DFT
Let us start from the usual non-relativistic electronic Hamiltonian
\begin{equation}\label{eq:ESham}
\hat{H}=\hat{T}+\hat{V}_{ee} + \hat{V},
\end{equation}
with $\hat{T}=-\sum_i^N\frac{\nabla_{i}^2}{2}$ the kinetic energy, $ N $ the number of particles in the system, $\hat{V}_{ee} $ the Coulomb interaction between all electron pairs, and $ \hat{V} = \sum_i^N v (i)$ the $N$-particle sum of the external potential, (typically) given by the positive field of the nuclei, felt by each electron.  
The lowest (ground) eigenstate associated with this Hamiltonian is labelled $\Psi$. 
According to the Hartree-Fock approximation, the expectation value of $\hat{H}$ is minimized in the space of Slater determinants. %, rather than in the whole space of anti-symmetric $N$-particle wavefunctions. 
Slater determinants are defined as $ \Phi  := \sum_{P}(-1)^P\psi_{P(1)}(\mathbf{x}_1)\cdots\psi_{P(N)}(\mathbf{x}_N)$, where the $\psi_n(\mathbf{x})$ are single-particle wave functions, spatial and spin coordinates are considered separable, i.e. $\psi_n(\mathbf{x}) \equiv \phi_n(\rv) s_{n}(\sigma)$, and the index $ P $ lists all possible permutations. The minimizer of this search is the so-called Hartree-Fock state:
\begin{equation}\label{eq:HFwf}
| \Phi^\hf \rangle =\argmin_{\Phi} \langle \Phi| \hat{H}|\Phi \rangle.
\end{equation}

Consider now the following $ \lam$-dependent Hamiltonian
\begin{equation}\label{eq:MPac}
\hat{H}_\lam^\hf = \hat{T} +\hat{V}_\hf +\hat{V} +\lam \left( \hat{V}_{ee}-\hat{V}_\hf\right)
\end{equation}
where $ \hat{V}_\hf =\sum_{i,j}^N \left(\hat{J}_j ^\hf(\bx_i) - \hat{K}_j^\hf(\bx_i) \right)  $, 
\begin{equation}\label{eq:Ji}
\hat{J}_i^\hf (\bx) = \int \frac{|\psi_i^\hf (\bx')|^2}{|\br -\br'|} \,\ud \bx'
\end{equation}
and $ \hat{K}_i^\hf $, which can be defined via its action on a test function $ \phi (\bx) $, reads
\begin{equation}\label{eq:Ki}
\hat{K}_i^\hf (\bx) \phi (\bx) = \psi_i^\hf(\bx) \int \frac{\psi_i^{\hf \ast}(\bx') \, \phi (\bx')}{|\br -\br'|}\, \ud \bx'.
\end{equation}
The $\lam$-dependent Hamiltonian in eq~\eqref{eq:MPac} is such that when $\lam=1$, we recover the interacting Hamiltonian [Eq.\eqref{eq:ESham}], while when $\lam=0$, we recover the HF Hamiltonian, $\hat{H}^\hf$. $\hat{H}^\hf$ is the Hamiltonian that has the HF wave function, $ \Phi^\hf $, as its ground state. 
For general $\lam$, the ground-state (GS) wavefunction of eq~\eqref{eq:MPac} is denoted $ \Psi_\lam^\hf $ and the GS energy $ E_\lam^\hf  $.

Defining the correlation energy for the HF reference as
\begin{equation}\label{eq:EcHF}
E_c^\hf = \langle \Psi | \hat{H} | \Psi \rangle-\langle \Phi^\hf | \hat{H} | \Phi^\hf  \rangle,
\end{equation}
the Hellmann-Feynman theorem on eq~\eqref{eq:MPac} yields
\begin{equation}
E_c^\hf = \int_0^1 W_{\lam}^\hf \ud \lam
\end{equation}
with
\begin{equation}\label{eq:Wlmp}
W_{\lam}^\hf := \langle \Psi_\lam^\hf |\hat{V}_{ee} - \hat{V}_\hf |\Psi_\lam^\hf  \rangle +  c_0^\hf [\ron^\hf].
\end{equation}
The constant shift $ c_0^\hf$ is equal to $U_\text{H}[\ron^\hf]+E_x[\{ \psi_i^\hf\}]$, where 
$U_\text{H}[\ron]=\frac{1}{2}\int \int \frac{\ron(\br) \ron(\br')}{|\br-\br'|} \ud \br \ud \br' $ is the mean field repulsion energy,  and $E_x[\{\psi_i \}] = -\frac{1}{2} \sum_{i, j}^{N} \int  \int  \frac{\psi_i^* (\bx) \psi_j^*(\bx) \psi_i (\bx') \psi_j (\bx')}{|\br -\br'|} \ud \bx \ud \bx'$ is the exchange energy, which comes from evaluating the interaction operator on a Slater determinant and subtracting the mean field term.

The notation $W_{\lam}^\hf $ has been adopted~\cite{SeiGiaVucFabGor-JCP-18} for the adiabatic connection integrand including an additional $E_x [\{\psi_i^\hf \}]$, which is however a $\lam$-independent quantity. The only difference between the two definitions is that when $\lam=0$, $W_{\lam}^\hf$ of eq~\eqref{eq:Wlmp} (which is elsewhere referred to as ``$W_{\lam, c}^\hf $'') is gauged to go to zero rather than to $E_x [\{\psi_i^\hf \}]$.

The small-$ \lam $ expansion of $W_{\lam}^\hf $ recovers the renowned M\o ller-Plesset series, i.e.
\begin{equation}\label{eq:MPseries}
W_{\lam\to 0}^\hf = \sum_{\text{n}=2}^\infty \text{n}\,E_c^{\text{MPn}} \lam^{\text{n}-1}.
\end{equation}
%While its expansion around the opposite limit, i.e. $ \lam \to \infty $, has been recently shown~\cite{SeiGiaVucFabGor-JCP-18, DaaGroVucMusKooSeiGieGor-JCP-20} to yield 
%\begin{equation}\label{eq:WlargeHF}
%W_{\lam\to \infty}^\hf  = W_\infty^\hf + W_\frac{1}{2}^\hf \lam^{-\frac{1}{2}} + W_\frac{3}{4}^\hf \lam^{-\frac{3}{4}} + o(\lam^{-\frac{3}{4}}).
%\end{equation}
As for its expansion around the opposite limit, in this work, we shall only be concerned with the leading-order term, $ W_\infty^\hf$, 
\begin{equation}\label{eq:WlargeHF}
W_{\lam\to \infty}^\hf  = W_\infty^\hf +o(\lam^0),
\end{equation}
whose explicit expression reads
\begin{equation}
W_\infty^\hf = E_{el}[\ron^\hf] + E_x[\{\psi_i^\hf\}],
\end{equation}
with
{\small
\begin{equation}\label{eq:EelHF}
	E_{\rm el}[\ron]\equiv \min_{\{\rv_1\dots\rv_N\}}\left\{\sum_{i,j>i}^{N}\frac{1}{|\rv_i-\rv_j|}-\sum_{i=1}^N v_{\rm H}(\rv_i;[\ron])+U_\text{H}[\ron]\right\},
\end{equation}}

\noindent the minimum total electrostatic energy of $N$ equal classical point charges $(-e)$ in a positive background with continuous charge density $(+e)\ron(\rv)$. 
%say something about the asymptotic wave function?
%introduce asymptotic Hamiltonian

We also recall that
\begin{eqnarray}\label{eq:PsiHFinfdef}
	\lim_{\lambda\to\infty}\Psi_\lambda^{\rm HF} & = & \mathop{\rm argmin}_\Psi\langle\Psi| \hat{H}_\infty^{\rm HF} |\Psi\rangle,
\end{eqnarray}
with $ \hat{H}_\infty^{\rm HF} :=  \hat{V}_{ee}-\sum_{i,j}^N\sum_{\sigma_i}\hat{J}_j^\hf (\bx_i) $~\cite{SeiGiaVucFabGor-JCP-18}.
Since $\hat{H}_\infty^{\rm HF}$ is a purely multiplicative operator, 
the square modulus  of its minimizing wave function, $|\Psi_\infty^{\rm HF}|^2$, is a classical distribution in ${\mathbb{R}}^{3N}$ localised where 
$\hat{H}_\infty^{\rm HF}$ as a function of $\rv_1,...,\rv_N$ attains its global minimum, i.e.,
\begin{equation}\label{eq:PsiHFinf}
|\Psi_\infty^{\rm HF}|^2=\frac{1}{N!}\sum_{\wp =1}^{N!}\prod_{i=1}^N \delta(\br_i-\br_{\wp (i)}^\text{min}).
\end{equation}
Equation~\eqref{eq:PsiHFinf} essentially tells us that the particles sit at fixed positions with respect to one another, forming a perfect crystal (with translational and rotational freedom), while the permutations $ \wp(i) $ account for the indistinguishability of the particles.
As mentioned in the Introduction, beside the interest that the $\lam \to \infty$ limit might raise \textit{per se},  $W_\infty^\hf$ is used in interpolation formulas that have proven worthwhile for the determination of properties of systems at their physical ($\lam =1$) interaction strength (such as for the determination of interaction energies of non-covalent complexes\cite{VucGorDelFab-JPCL-18}). 

Let us now review the theory of the adiabatic connection typically considered in DFT: the density-fixed adiabatic connection.\cite{HarJon-JPF-74, GunLun-PRB-76, LanPer-SSC-75, Lan-PRL-84}

Consider the Levy-Lieb $\lambda$-dependent functional~\cite{Lev-PNAS-79}
\begin{equation}
F_{\lambda}[\ron]:= \min_{\Psi\to\ron}\langle \Psi | \hat{T} + \lambda \hat{V}_{ee} | \Psi \rangle,
\label{eq:Flambda}
\end{equation}
with $\Psi_{\lambda}^\ks[\ron]$ the minimizer of the above search.
 Assuming that $\ron$ is $v$-representable for all $\lambda$, one can write the following $\lambda$-dependent Hamiltonian:
 \begin{equation}\label{eq:Hldft}
\hat{H}_{\lambda}^\ks= \hat{T} + \lambda \hat{V}_{ee} + \hat{V}^{\lambda},
\end{equation}
where $\hat{V}^{\lambda}=\sum_i^N v^{\lambda}(\br_i)$, and
\begin{equation}\label{eq:vextlambda}
v ^\lambda [\ron_0] (\br) = -\frac{\delta F_{\lambda}[\ron]}{\delta \ron}\Big|_{\ron=\ron_0}  (\br)
\end{equation}
is the local external potential that enforces the prescribed density $\ron$ at each $\lam$.
Defining the XC energy of KS-DFT as
\begin{equation}\label{eq:ExcKS}
 E_{xc}^\ks [\ron] =F_1[\ron] - F_0[\ron] - U_\text{H}[\ron],
\end{equation}
the Hellmann-Feynman theorem on eq~\eqref{eq:Flambda} yields
 \begin{equation}
 E_{xc}^\ks [\ron] = \int_0^1 W_\lam^\ks [\ron] \ud \lam
 \end{equation}
with
\begin{equation}\label{eq:Wldft}
W_\lam^\ks [\ron] := \langle \Psi_{\lambda}^\ks[\ron] | \hat{V}_{ee}|\Psi_{\lambda}^\ks[\ron]\rangle-U_\text{H}[\ron].
\end{equation}
A small note is that, in some sense, in DFT it is more natural to define the XC energy rather than only the correlation part. However, the exchange energy of DFT is formally analogous to that of HF, in the cases in which the KS state $\Psi_0^\ks $ is a Slater determinant, $\Phi^\ks$, something which is usually assumed.
Then the correlation energy in DFT can be defined as
\begin{equation}\label{eq:EcKS}
 E_c^\ks [\ron] = E_{xc}^\ks [\ron] - E_x[\{\psi_i^\ks\}[\ron]],
\end{equation}
where the $\psi_i^\ks$ are the orbitals that form the KS determinant.

The most conspicuous difference between the adiabatic connection formalism introduced in eq~\eqref{eq:MPac} and the DFT adiabatic connection is that, in this latter, the density is kept fixed along $\lam$. In other words, there is a subtle dependence on $ \lam $ sneaking in via the $ \hat{V}^\lam $ operator. As a consequence, while the MP AC integrand corresponds to the derivative of the $\lam$-dependent energy (minus a shift), i.e.,
\begin{equation}\label{eq:WlasEp}
W_\lam^\hf [\ron^\hf]= \frac{\ud}{\ud \lam}E_\lam^\hf [\ron^\hf]+ c_0^\hf[\ron^\hf],
\end{equation}
$W_\lam^\ks $ is rather the derivative of $ F_{\lambda} $ (minus a shift):
\begin{equation}
W_\lam^\ks [\ron] = \frac{\ud}{\ud \lam} F_{\lambda}[\ron] + c_0^\ks,
\end{equation}
with $ c_0^\ks = - U_\text{H}[\ron] $.
In turn, $F_{\lambda}$ is also equal to $E_\lam^\ks[\ron]-V_\lam[\ron] $, with $ E_\lam^\ks  $ the GS energy of Hamiltonian~\eqref{eq:Hldft} and $ V_\lam[\ron] := \langle  \Psi_{\lambda}^\ks[\ron] | \hat{V}^{\lambda}|\Psi_{\lambda}^\ks[\ron]\rangle  $.
By reshuffling eq~\eqref{eq:Hldft}, one realizes that the fluctuation or perturbation potential -- that is, the operator which is turned on by $\lam$ -- has the form $\left(\hat{V}_{ee}+\frac{\hat{V}^\lam - \hat{V}^{\lam=0}}{\lam}\right)$. 
Then, the small-$\lam$ expansion of $ W_\lam^\ks $,~\cite{GorLev-PRB-93, GorLev-PRA-94}
\begin{equation}\label{eq:GLseries}
W_{\lam\to 0}^\ks [\ron] =E_x [\{\psi_i^\ks [\ron] \}]+ \sum_{\text{n}=2}^\infty \text{n}\,E_c^{\text{GLn}} \lam^{\text{n}-1},
\end{equation}
contains also the order-by-order expansion of the fluctuation potential inside the perturbation series coefficients $ E_c^{\text{GLn}}  $, adding a layer of complexity to the usual MP expressions.
The $\text{n}=2$ term reads
\begin{equation}
E_c^\text{GL2} [\ron] =\sum_{i=1}^\infty \frac{|\langle \Phi^\ks|\hat{V}_{ee} -\sum_{j=1}^N v_{\text{H}x} (\br_j) |\Phi^\ks_i \rangle|^2}{E^\ks_0 -E^\ks_{0,i}},
\end{equation}
where $ v_{\text{H}x} = v_\text{H} + v_x $  with $  v_{x} = \frac{\delta \, E_x[\{\psi_i^\ks [\ron] \}]}{\delta \ron}  \Big|_{\ron=\ron_0}$ and $ \Phi^\ks_i $ and $ E^\ks_{0,i} $ are the excited KS states and energies.
The explicit computation of any subsequent term seems absent from the literature.

The leading order of the large-$\lambda$ expansion of $W_{\lambda}^\ks$ is also a constant,\cite{GorVigSei-JCTC-09} analogously to eq~\eqref{eq:WlargeHF}:
\begin{equation}
\label{eq:lambdainfDFT} 
W_{\lambda\rightarrow\infty}^\ks[\ron] = W_\infty^\ks [\ron] +  o\left( \lam^{-\frac{1}{2}} \right).
\end{equation}
%from T.Daas paper : "The presence of the order λ ^−3/4 is interesting because this term is zero in the large λ-expansion of the DFT adiabatic connection.[33]"

%SCE wave function
The asymptotic wavefunction in the DFT adiabatic connection is defined as
\begin{equation}\label{eq:DFTwfinf}
\Psi_\infty^\ks = \text{argmin}_{\Psi \to \ron} \langle \Psi |\hat{H}_\infty^\ks | \Psi \rangle,
\end{equation}
with $ \hat{H}_\infty^\ks =\hat{V}_{ee} + \hat{V}^\infty $
and where $\hat{V}^\infty $ is the $ N$-electron sum of the one-body operator, 
\begin{equation}\label{eq:vinfDFT}
v^\infty [\ron_0](\br)=-\frac{\delta F_\infty[\ron]}{\delta n}\Big|_{\ron=\ron_0},
\end{equation}
with $ F_\infty[\ron]:= \lim_{\lam \to \infty} \frac{F_\lam [\ron]}{\lam}$.
In a way reminiscent of eq~\eqref{eq:PsiHFinf} but heavily complicated by the density constraint, $ \big| \Psi_\infty^\ks\big|^2 $ is a semi-classical distribution
\begin{eqnarray}\label{eq:PsiSCE}
\big| \Psi_\infty^\ks\big|^2 =\frac{1}{N!}\sum_{\wp =1}^{N!}\int d\textbf{s}\frac{\ron(\textbf{s})}{N}\prod_{i=1}^N \delta\left(\textbf{r}_i -\textbf{f}_{\wp (i)}(\textbf{s})\right),
\end{eqnarray}
where the co-motion functions $ \textbf{f}_i $ are %extremely involved 
mathematical objects which parameterize the set of all configurations where $\hat{H}_\infty^\ks  $ is minimum. There are $N-1$ non-trivial co-motion functions which provide the position of $N-1$ particles, given the position of a reference one.
%The minimum of $\hat{H}_\infty^\ks  $ has to be degenerate on a \textit{continuum} of dimensionality at least equal to the domain of the density. 

Note the difference between Eqs.~\eqref{eq:PsiHFinf} and~\eqref{eq:PsiSCE} : whereas $\big| \Psi_\infty^\hf\big|^2$ is a perfect crystal with well defined positions, $\big| \Psi_\infty^\ks\big|^2$ is rather a superposition of infinitely many cristals, since the $N-1$ particle positions depend parametrically on the position of a single one which varies freely (for a more focused description of the strong-interaction limit of DFT, the interested reader is referred to references~\onlinecite{Sei-PRA-99, SeiGorSav-PRA-07, GorVigSei-JCTC-09,GiaVucGor-JCTC-18, Lew-CRM-18}).
 
Finally, we recall that, using the Legendre transform formulation of Lieb,\cite{Lie-IJQC-83} it has been shown~\cite{SeiGiaVucFabGor-JCP-18} that, for a given density,
\begin{equation}\label{eq:ineqWinf}
W_\infty^\hf [\ron] \leq W_\infty^\ks [\ron].
\end{equation}

Throughout this work, we will use the superscript ``SD'' to indicate both the KS and the HF versions of a given quantity. For example, with $E_c^\text{SD}$, we mean $E_c^\hf$ and/or $E_c^\ks$.

\subsection{The Hubbard Dimer}\label{sec:HD}
The Hubbard model is often used to test new methods and concepts in chemistry and physics because its Hamiltonian is extremely simple compared to physical systems (atoms and molecules), while still incorporating many of the correlation effects in such systems. Its two-site version, considered in this work, reads:
\begin{equation}\label{eq:HDham}
\mathcal{\hat{H}}=\mathcal{\hat{T}}+\mathcal{\hat{U}}+\mathcal{\hat{V}}
\end{equation}
where
\begin{eqnarray}
\mathcal{\hat{T}}&=& -t \sum_\sigma \left(\hat{a}_{0\sigma}^\dagger\hat{a}_{1\sigma}+\hat{a}_{1\sigma}^\dagger\hat{a}_{0\sigma}\right)\label{eq:Tdef}\\
\mathcal{\hat{U}}&=&U \sum_{i=0,1} \hat{n}_{i \uparrow}\hat{n}_{i \downarrow}  \label{eq:Uhat}\\
\mathcal{\hat{V}}&=&\sum_{i=0,1} v_i \hat{n}_i,
\end{eqnarray}
 $\hat{a}^\dagger,\, \hat{a} $ are the usual creation and annihilation operators, 
 $\sigma= \uparrow, \downarrow$ labels the spin of the particles, $i=0,1 $ labels the two sites, and $\hat{n}_{i\sigma} = \hat{a}_{i\sigma}^\dagger\hat{a}_{i\sigma}$ and $\hat{n}_i= \hat{n}_{i\sigma}+\hat{n}_{i\overline{\sigma}}$ (with $\overline{\sigma}$ being the spin opposite to $ \sigma $) are the occupation operators.
 The reduced variables $u=\frac{U}{2\,t}$  and $\delta v=\frac{\Delta v}{2\, t}$, with $\Delta v = v_1-v_0 $, fully determine the eigenstates of the Hamiltonian~\eqref{eq:HDham}. Given this, the hopping parameter $t$ is set to $1/2$ throughout the paper and the gauge, i.e. $c=v_0+v_1$, to zero, as is customary.~\citep{CarFerSmiBur-JPCM-15} Furthermore, we consider the dimer at ``half-filling,'' which means that the sum of the expectation value of the occupation operators on each site is set to two (i.e., $  n_0+n_1= 2 $) and restrict ourselves to singlet states ($S_z =0$). The site occupation difference $\Delta n$ corresponds to the electron probability density in the model and is defined as the expectation value of the difference between the site-occupation operators, $\Delta n =\langle \Psi | \hat{n}_1-\hat{n}_0 |\Psi\rangle $.
The three-dimensional  Hilbert space is represented in the basis  $ | 0\uparrow 0\downarrow\rangle,|1\uparrow1\downarrow \rangle $, and $ \frac{1}{\sqrt{2}} \left( | 0\uparrow 1\downarrow\rangle-|0\downarrow 1\uparrow\rangle\right)$.  
 Note that all energy terms are symmetric with respect to the change in sign of $\Delta v$, while the sign of the site-occupation difference is opposite to that of the external potential difference. We call $\epsilon (U, \Delta v)$ the ground-state energy associated with Hamiltonian~\eqref{eq:HDham}.

The restricted Hartree-Fock Hamiltonian for this model can be written as:~\citep{GiaPri-JCP-22}
\begin{equation}\label{eq:HFhamTB}
\mathcal{\hat{H}}^\text{RHF} = \mathcal{\hat{T}}+ \text{$\mathcal{\hat{\tilde{V}}}$},
\end{equation}
with
$\mathcal{\hat{\tilde{V}}} =\sum_{i=0,1}  \tilde{v}_i \hat{n}_i$ and $\tilde{v}_i = v_i + U\frac{n_i^\hf}{2}$. 
The symbol $n_i^\hf$ indicates the HF site occupation on each site  (as converged to its stationary point), and $U \frac{n_i^\hf}{2}$ is the site mean field potential.
Note that, because $n_0^\hf+n_1^\hf = 2 $, setting $ v_0 +v_1=0 $ in eq~\eqref{eq:HDham} forces the sum of $\tilde{v}_i$ to give $\tilde{v}_0+\tilde{v}_1 = U$.

\section{MP and DFT adiabatic connections for the Hubbard Dimer}\label{sec:ACforHD}
Using Eqs.~\eqref{eq:HDham} and~\eqref{eq:HFhamTB}, the M\o ller-Plesset adiabatic connection [eq~\eqref{eq:MPac}] for the Hubbard dimer reads
\begin{small}
\begin{eqnarray}\label{eq:ACHFham}
\hat{H}_\lam^\hf & = & \hat{\mathcal{T}}  + \sum_i \tilde{v}_i \hat{n}_i +  \lam\, U\,  \sum_i \left( \hat{n}_{i\uparrow} \hat{n}_{i\downarrow} -\frac{n_i^\hf}{2}\hat{n}_i\right) \nonumber \\
& = & \hat{\mathcal{T}}  +  \lam\, U\, \sum_i \hat{n}_{i\uparrow} \hat{n}_{i\downarrow} + \sum_i v_i^{\lam,\hf} \hat{n}_i ,
\end{eqnarray}
\end{small}

\noindent where in the second line we have introduced the $ \lam$-dependent external potential, $ v_i^{\lam,\hf} $, defined as
\begin{equation}
 v_i^{\lam,\hf} := \tilde{v}_i - \lam\, U\, \frac{n_i^\hf}{2},
\end{equation}
and where the gauge -- defined as $ c^{\lam,\hf}  = v_0^{\lam,\hf} +v_1^{\lam,\hf} $ -- depends linearly on $ \lam $:
\begin{equation}
c^{\lam,\hf} =U \left(1-\lam \right).
\end{equation}
The associated $ \lam$-dependent ground state, $ \Psi_\lam^\hf, $ can be calculated explicitly for any pair of interaction parameter and external potential, $ \{ U, \Delta v\} $. 
 Consequently, it is also possible to study the $\lambda$-dependent behaviour of relevant quantities such as the adiabatic connection integrand, $W^\hf_\lam$ (section~\ref{sec:lACI}), and the site occupation difference, $ \Delta n^\hf_\lam $ (section~\ref{sec:lSOD}), analytically as functions of the variables $ \{ U, \Delta v\} $. They can also be expressed analytically as functions of $\Delta n^\hf$, the HF site occupation difference, in place of $\Delta v$, as the function  $f: \Delta v \to \Delta n^\hf$ is analytically invertible.~\citep{GiaPri-JCP-22}

The MP adiabatic connection integrand for our model system reads
{\small
\begin{equation}\label{eq:wlHD}
W_\lam^\hf = \langle \Psi_\lam^\hf |\mathcal{\hat{U}} -U\! \sum_i  \frac{n_i^\hf}{2} \hat{n}_i|  \Psi_\lam^\hf \rangle +\frac{U}{2}\left( 1 + \left( \frac{\Delta n^\hf}{2}\right)^2\right),
\end{equation}}

\noindent where the term in the Dirac brakets is simply $\frac{\ud \, E_\lam^\hf}{\ud \lam}$ and the remainder is $ c_0^\hf $ for the Hubbard dimer, i.e. the shift which makes $ W_0^\hf = 0$. %as in eq~\eqref{eq:Wlmp}.%

As for the density-fixed adiabatic connection, even in the  simple setting of the asymmetric Hubbard dimer, the $\lam$-dependent potential that keeps the density fixed cannot be determined in closed form (with the exception of the symmetric case, $\Delta n=0$~\cite{Fro-MP-15}). However, it can be computed quite efficiently by using Lieb's formulation,~\cite{Lie-IJQC-83}%. In the Hubbard dimer, the Lieb's convex formulation simply becomes~\cite{GiaLoo-arxiv-23}
\begin{equation}\label{eq:Vlambda}
\Delta v^{\lam, \ks} (U,\,  \Delta n) = \argmax_{\Delta v}\left( \epsilon \, (\lam\, U,\, \Delta v) - \frac{\Delta v}{2} \Delta n \right), 
\end{equation}
with $\Delta v^{\lam, \ks} =v^{\lam, \ks}_1-v^{\lam, \ks}_0$ and the usual gauge $v^{\lam, \ks}_1+v^{\lam, \ks}_0 = 0$. When $\lam=0$, $\Delta v^{\lam, \ks}$ is usually referred to simply as $\Delta v_s$.~\cite{GiaPri-JCP-22, CarFerSmiBur-JPCM-15}

 With $\Delta v^{\lam, \ks} $, we construct the $\lam$-dependent Hamiltonian of the density-fixed adiabatic connection [eq~\eqref{eq:Hldft}] for the Hubbard dimer,
\begin{equation}\label{eq:HlDFT}
H^\ks_\lam =\hat{\mathcal{T}}   +  \lam\, U\,  \sum_i  \hat{n}_{i\uparrow} \hat{n}_{i\downarrow} + \frac{\Delta v^{\lam, \ks} (U,\,  \Delta n) }{2} \left(\hat{n}_1-\hat{n}_0\right),
\end{equation}
and the corresponding AC integrand [eq~\eqref{eq:Wldft}],
\begin{equation} \label{eq:wlDFTHD}
W_\lam^\ks = \langle \Psi_\lam^\ks |\,\mathcal{\hat{U}} \,|  \Psi_\lam^\ks \rangle +\frac{U}{2}\left( 1 + \left( \frac{\Delta n}{2}\right)^2\right),
\end{equation}
where $ \Psi_\lam^\ks $ is the $\lam$-dependent ground state associated with the Hamiltonian given in eq~\eqref{eq:HlDFT}.
Differently than for eq~\eqref{eq:wlHD}, the term in Dirac brackets in~\eqref{eq:wlDFTHD} is not the derivative of the total energy with respect to $\lam$, but rather the derivative of $F^\lam $ [eq~\eqref{eq:Flambda}], which in this setting reads
\begin{equation}\label{eq:FlHD}
F^\lam (U,\, \Delta n) =\langle \Psi_\lam^\ks |\,\hat{\mathcal{T}}   +  \lam\, U\,  \sum_i  \hat{n}_{i\uparrow} \hat{n}_{i\downarrow}\,|  \Psi_\lam^\ks \rangle. 
\end{equation}
Equation~\eqref{eq:vextlambda} then translates into
\begin{equation}
\frac{\Delta v^{\lam, \ks} (U,\,  \Delta n) }{2}= - \, \frac{\partial F^\lam (U,\, \Delta n)}{\partial \Delta n}.
\end{equation}

Both AC integrands are defined in such a way that
\begin{equation}
E_c^\text{SD} = \int_0^1 W_\lam^\text{SD} \mathrm{d}\lam 
\end{equation}
with SD=HF, KS and $E_c^\text{SD} = \langle \Psi | \hat{\mathcal{H}} | \Psi \rangle-\langle \Psi_0^\text{SD} | \hat{\mathcal{H}} | \Psi_0^\text{SD} \rangle$ (compare Eqs.\eqref{eq:EcHF} and \eqref{eq:ExcKS},  \eqref{eq:EcKS}).

Introducing a generalized $\lam$-dependent correlation energy (not unusual in DFT~\cite{GorLev-PRB-93,RvL94}) as
\begin{equation}
E_{c, \lam}^\text{SD}=\int_0^\lam W_{\lam'}^\text{SD} \mathrm{d}\lam', 
\end{equation}
with $E_{c, 1}^\text{SD}=E_c^\text{SD}$, one finds, in the Hubbard dimer setting, that
\begin{equation}\label{eq:scalecorr}
E_{c, \lam}^\text{SD} (U, \Delta n) =E_c^\text{SD}(\lam \,U, \Delta n)
\end{equation}
for both types of correlation energies.

\section{Results}\label{sec:Results}
In this section, we present the analytical and numerical results obtained for the MP and DFT adiabatic connections, respectively.

\subsection{Shapes of the two adiabatic connection integrands}\label{sec:lACI}
\begin{figure*}
\centering
 \begin{tabular}[c]{cc}
	 {\begin{subfigure}{0.5\textwidth}
      \includegraphics[scale=0.4]{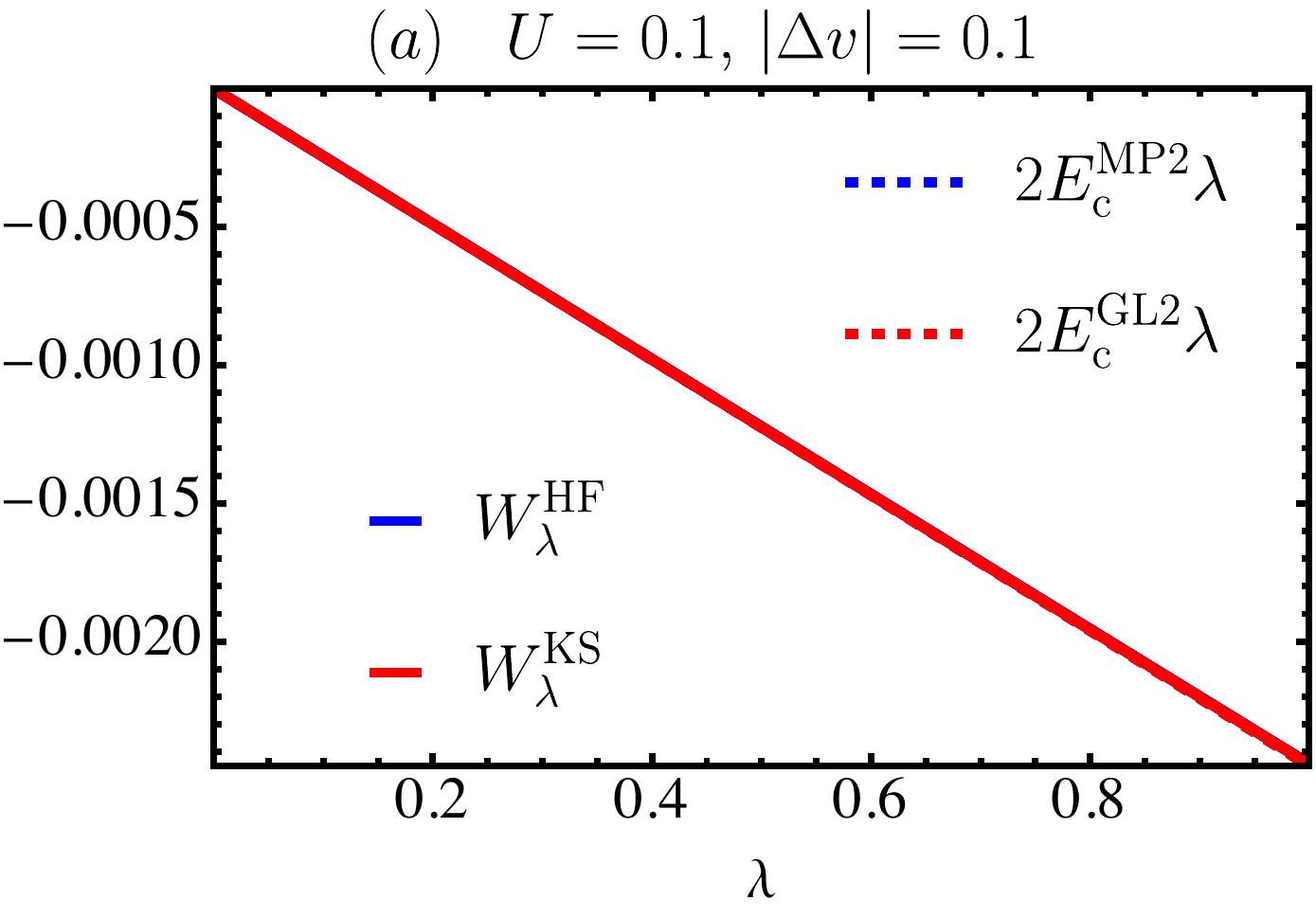}
    \end{subfigure}\phantom{O}
} & {\begin{subfigure}{0.5\textwidth}
 \includegraphics[scale=0.4]{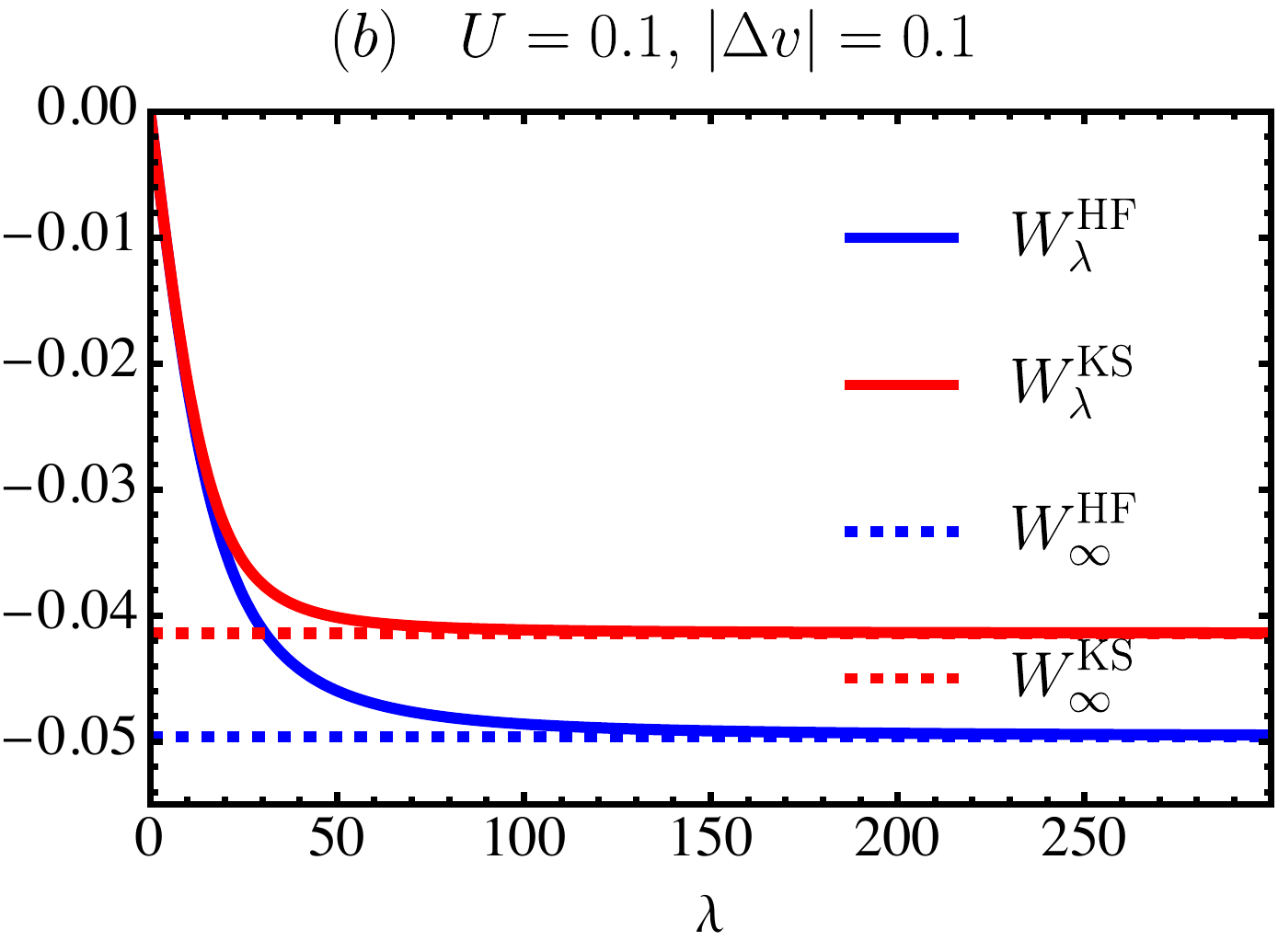}
 \end{subfigure}}\\
  {\begin{subfigure}{0.5\textwidth}
 \includegraphics[scale=0.4]{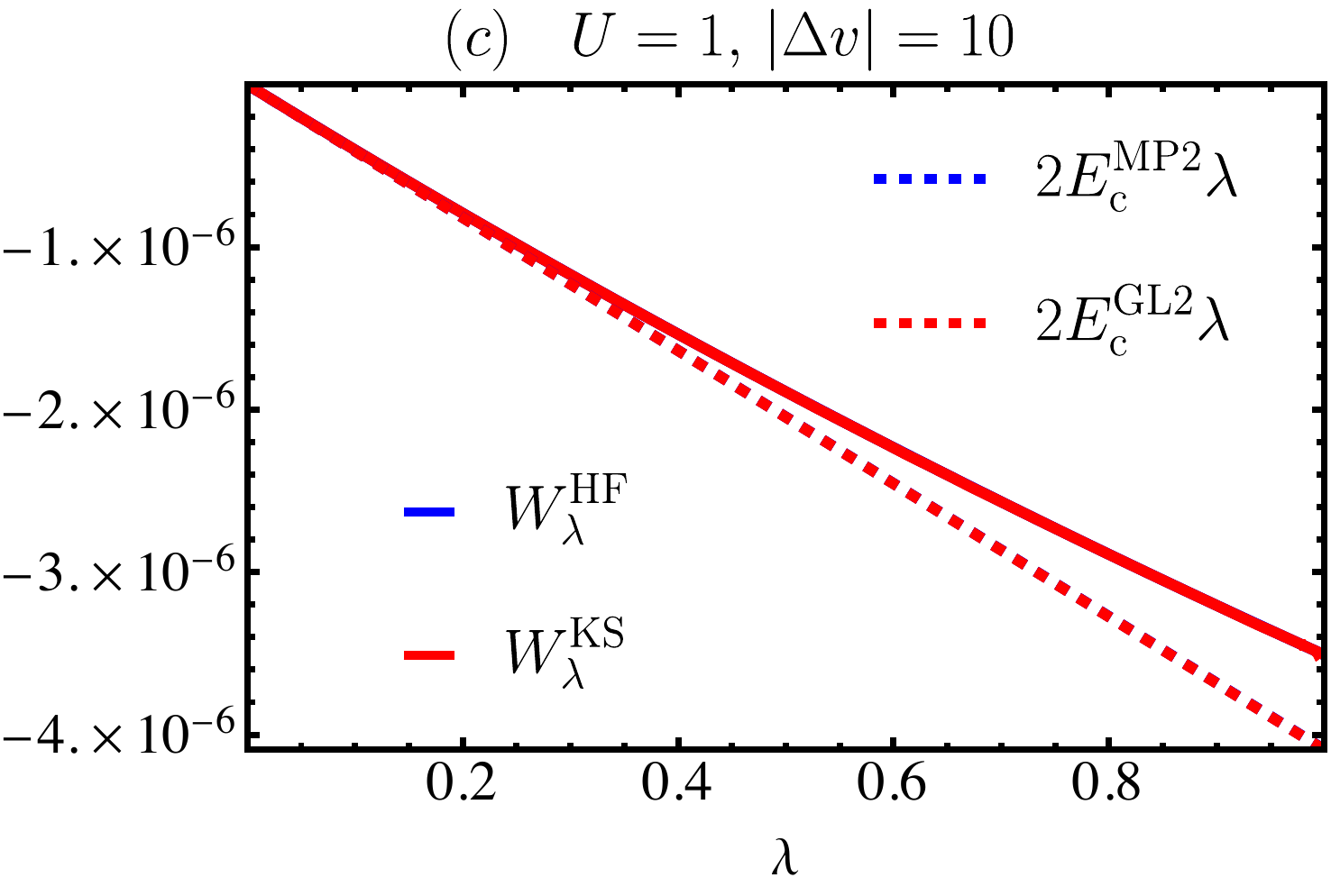}%
 \end{subfigure}\phantom{oo}}
 & {\begin{subfigure}{0.5\textwidth}
 \includegraphics[scale=0.41]{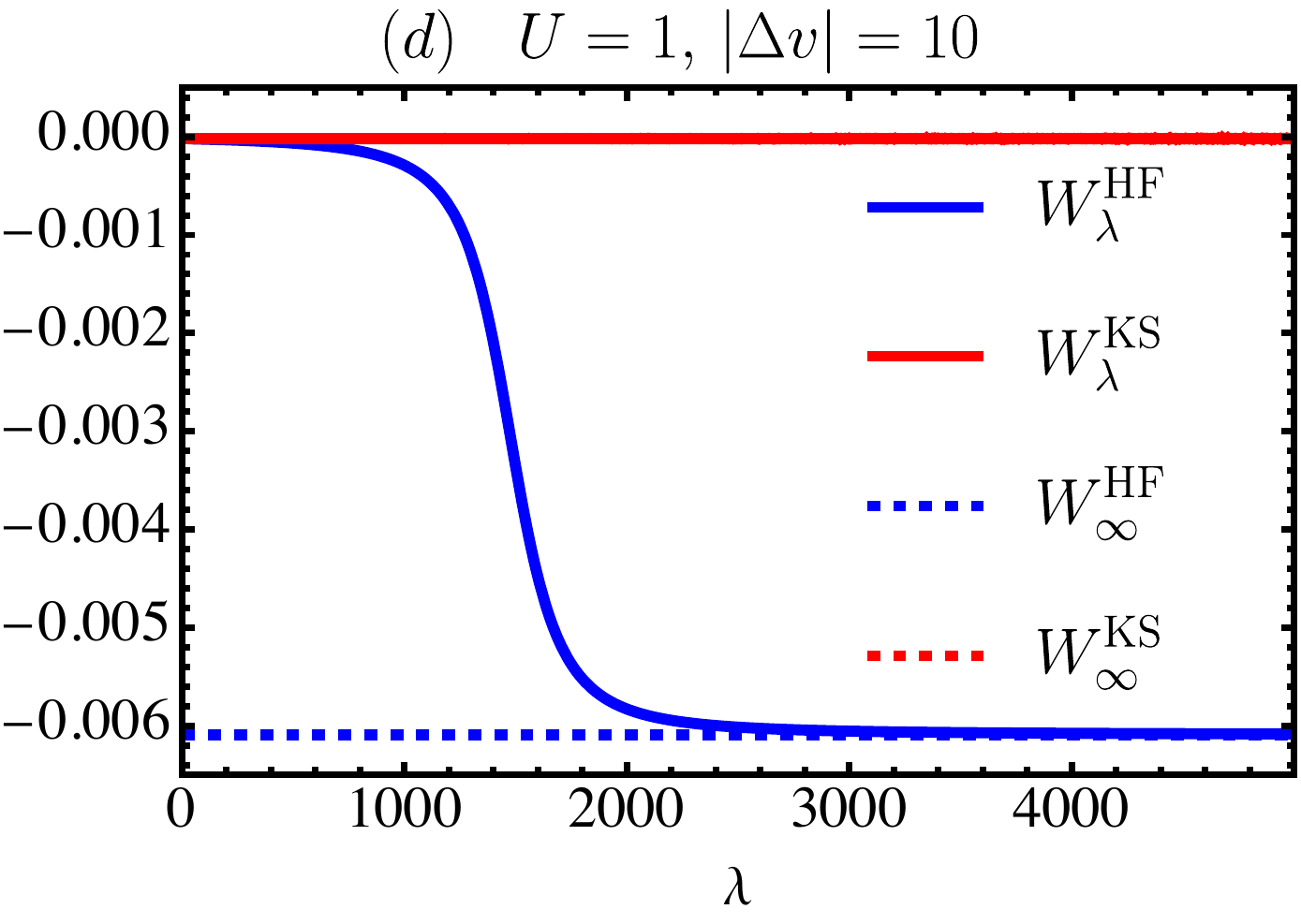}%
 \end{subfigure}}\\
 {\begin{subfigure}{0.5\textwidth}
 \phantom{o}     \includegraphics[scale=0.4]{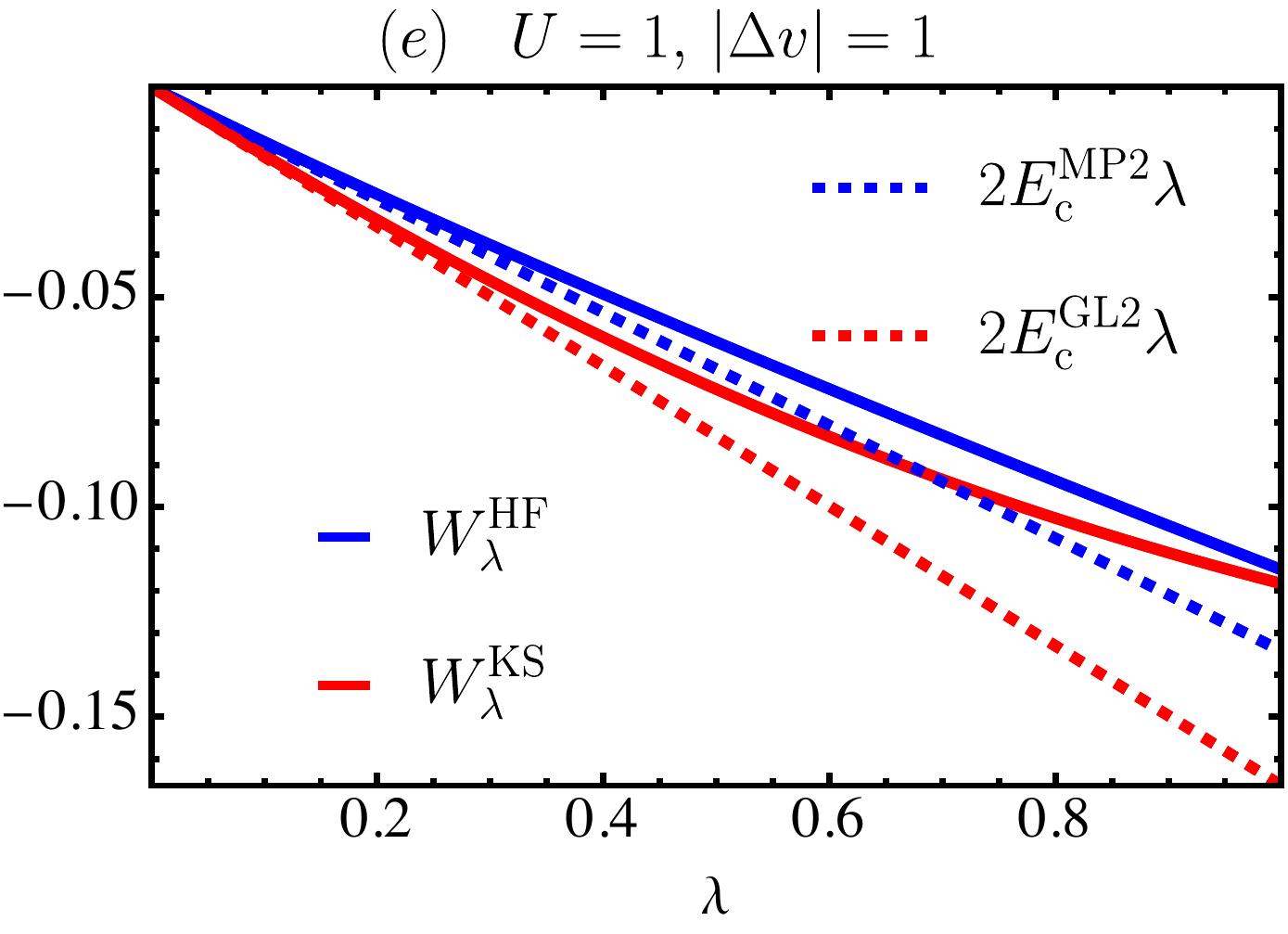}
    \end{subfigure}
} & {\begin{subfigure}{0.5\textwidth}
 \includegraphics[scale=0.4]{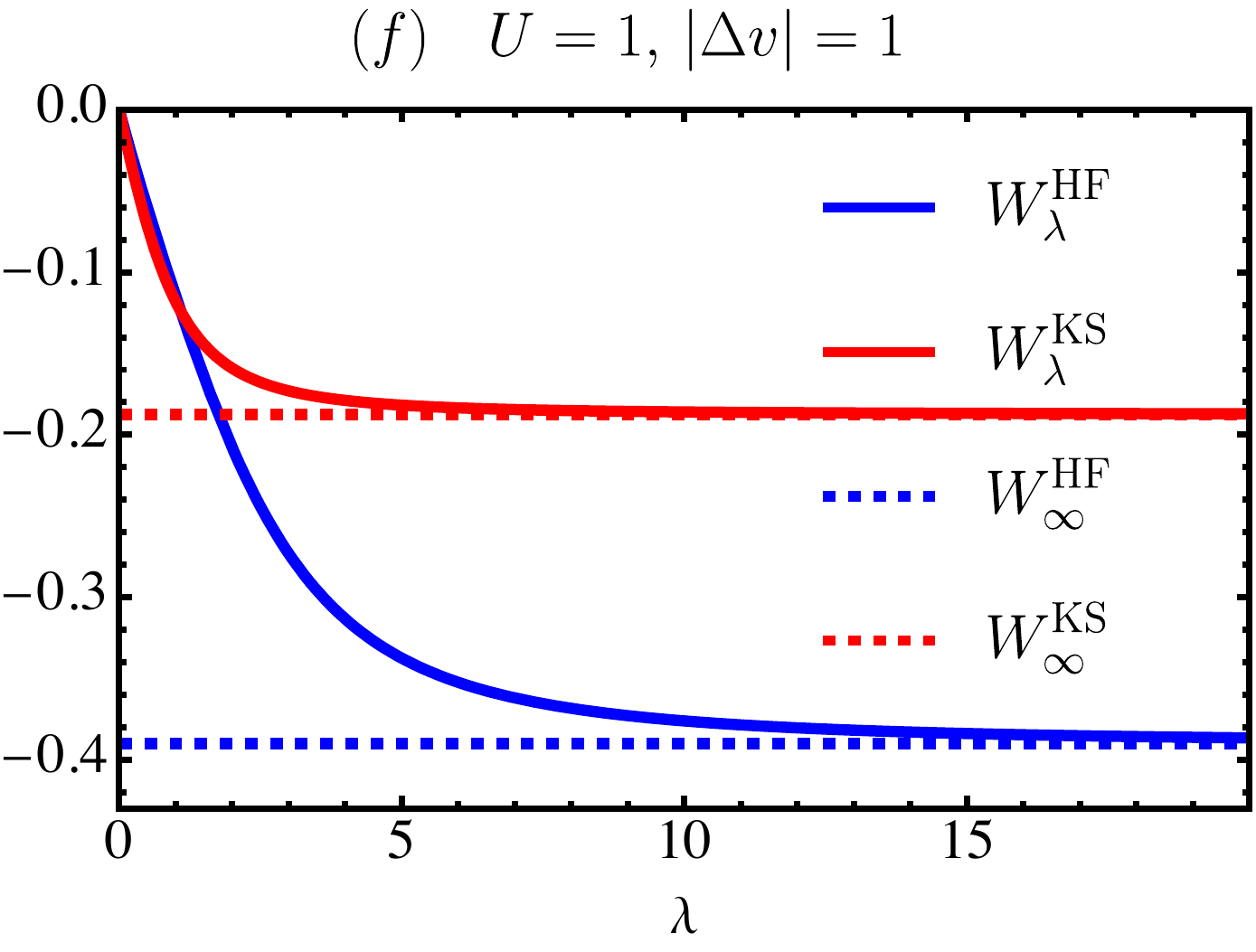}
 \end{subfigure}}\\
  {\begin{subfigure}{0.5\textwidth}
 \includegraphics[scale=0.4]{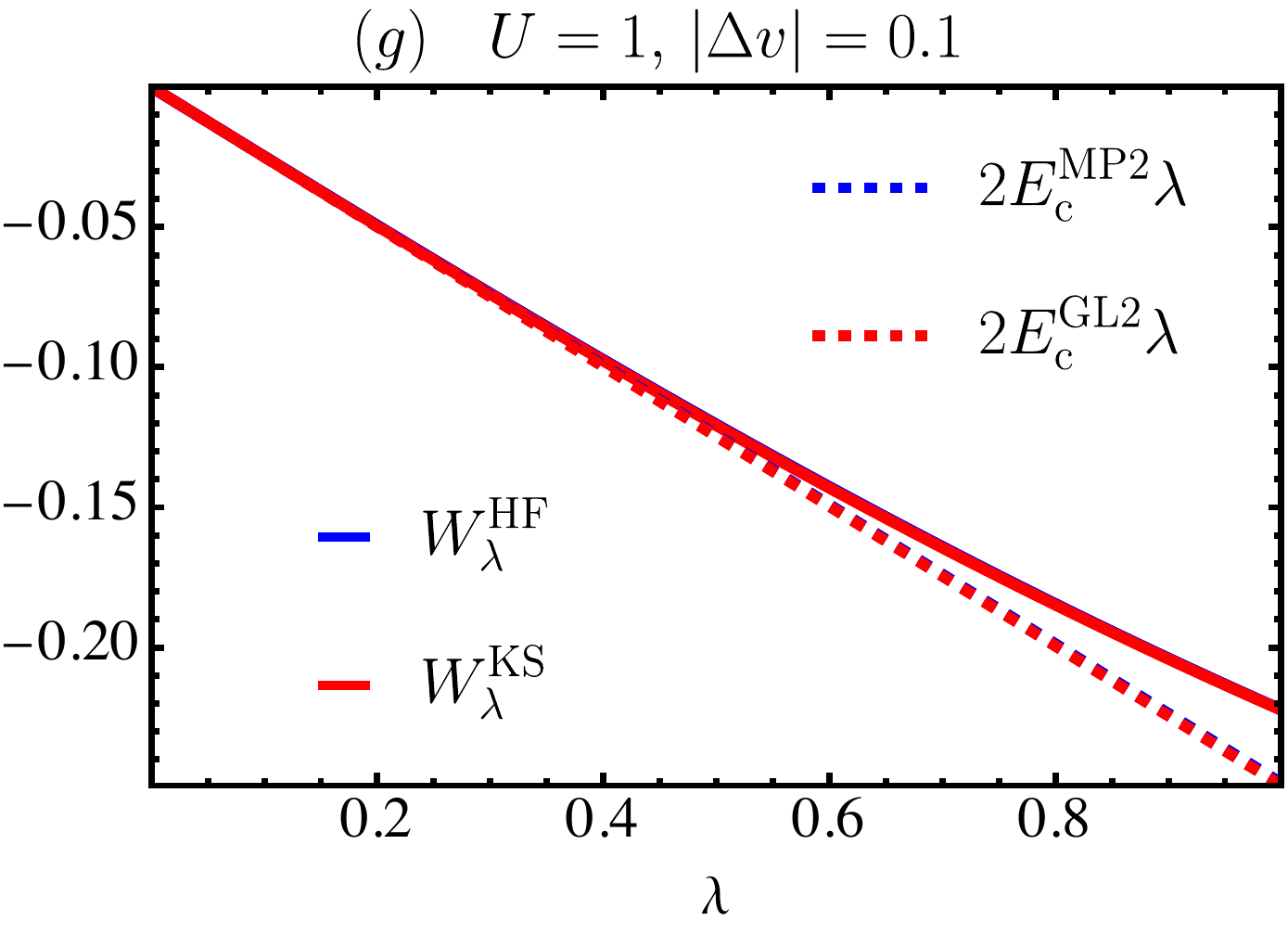}
 \end{subfigure}}
 & {\begin{subfigure}{0.5\textwidth}
 \includegraphics[scale=0.4]{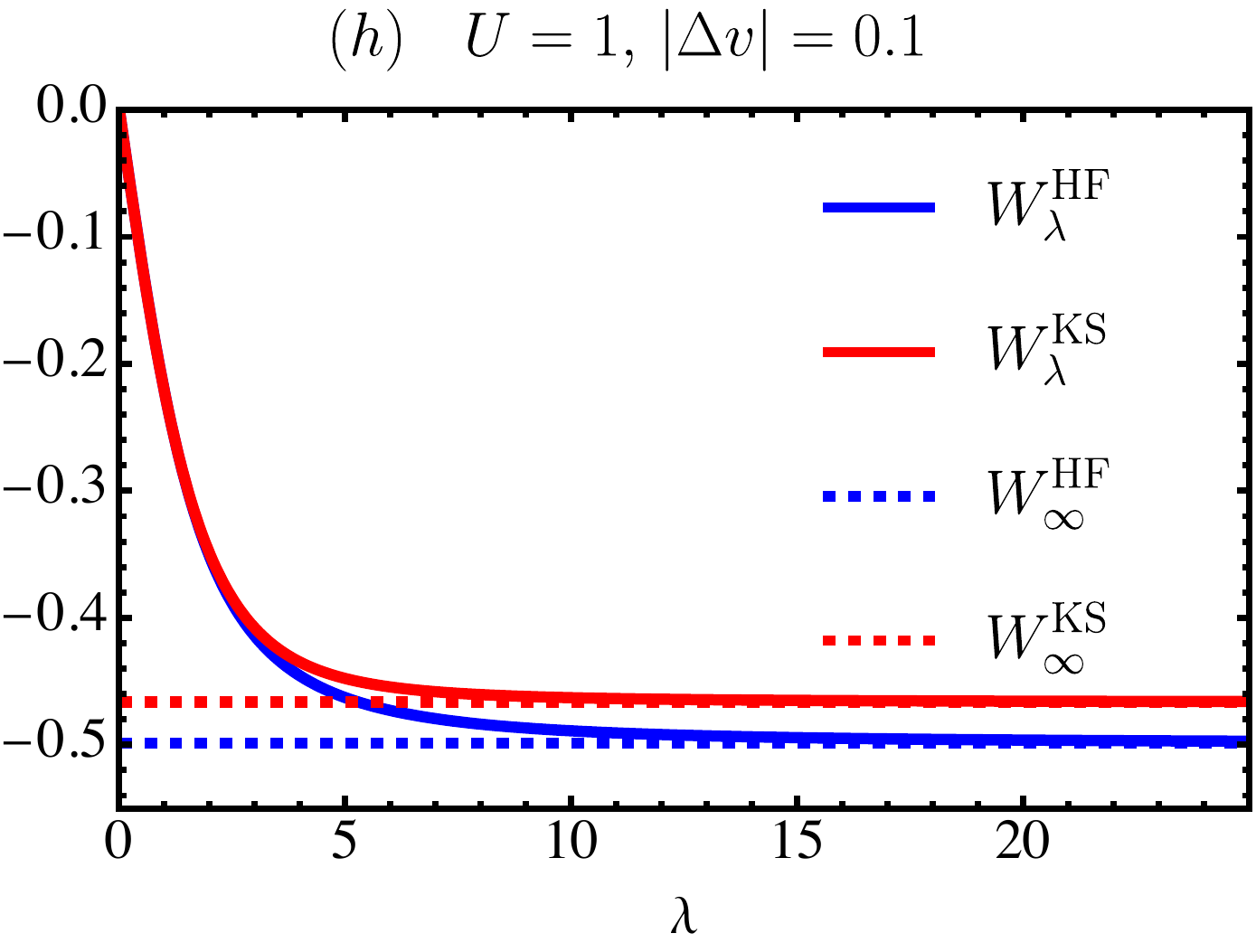}
  \end{subfigure}}\\
 {\phantom{oo}\begin{subfigure}{0.5\textwidth}
 \includegraphics[scale=0.38]{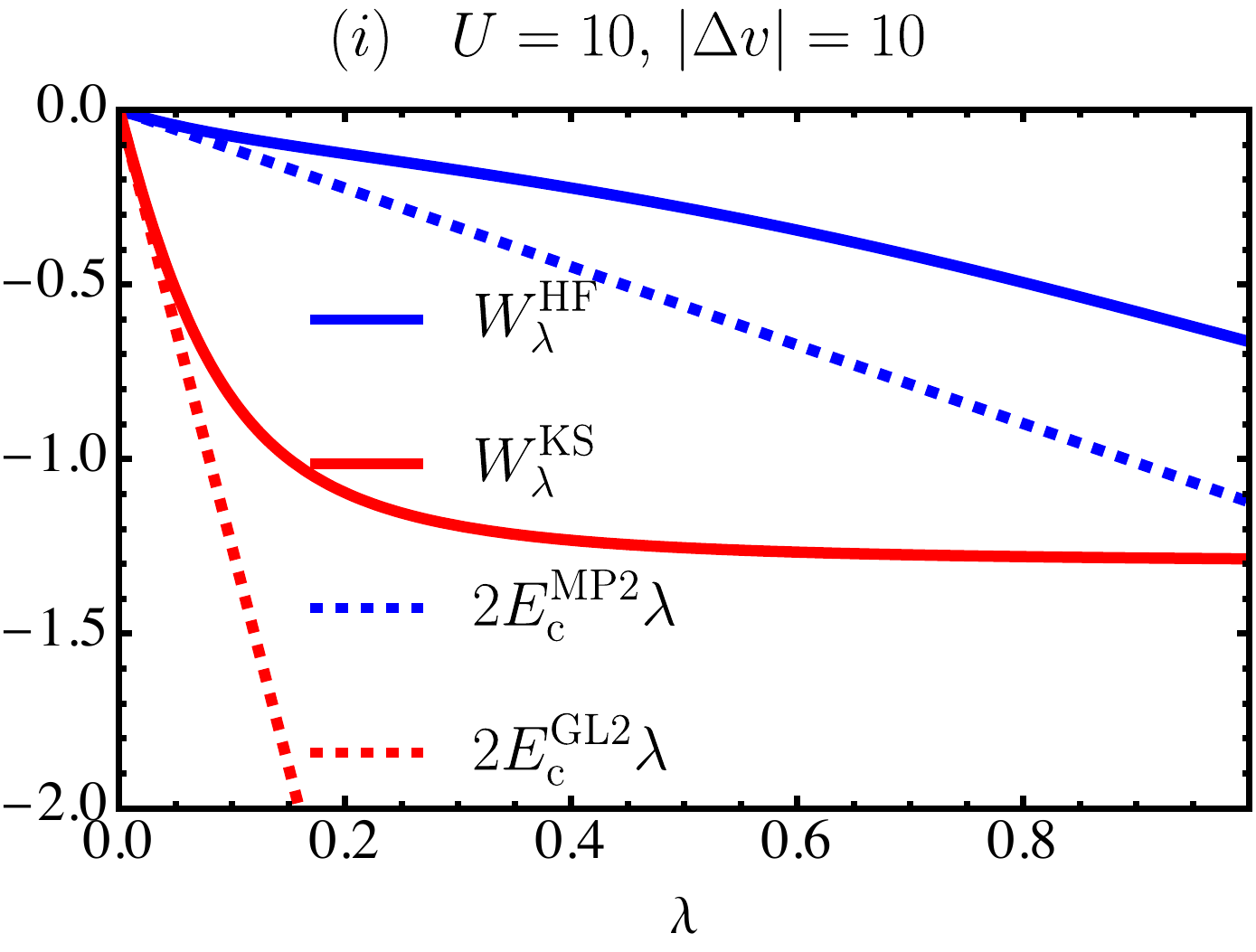}%ins
 \end{subfigure}}
 & {\begin{subfigure}{0.5\textwidth}
 \includegraphics[scale=0.4]{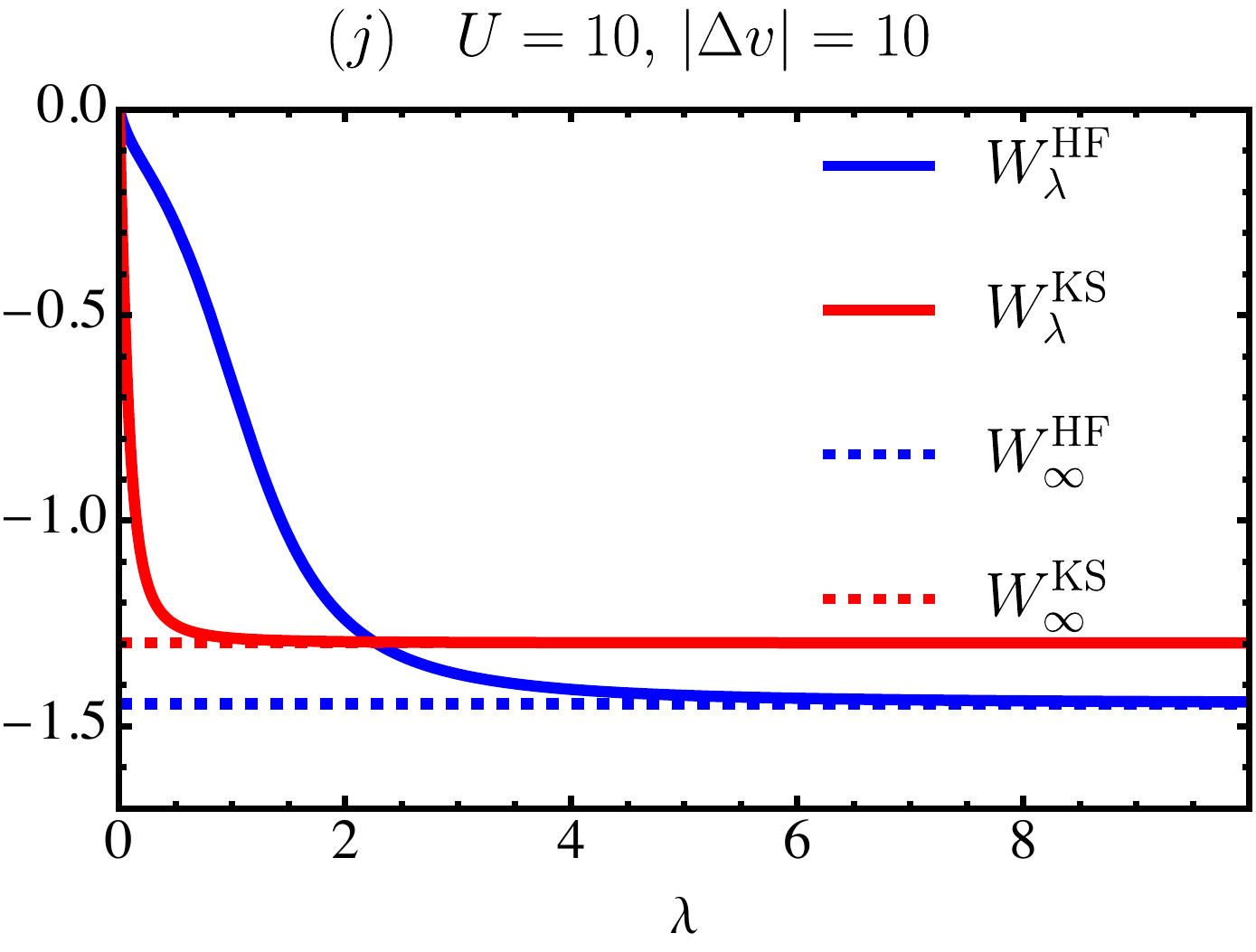}
 \end{subfigure}}
 \end{tabular}   
 \caption{\label{fig:Wlambda} %\small
 \footnotesize{Shapes of the two AC integrands, $W_\lam^\ks$ (solid red) and $W_\lam^\hf$ (solid blue), at different correlation regimes. In the left column, $\lam$ ranges between 0 and 1 and  $2\, E_\text{c}^\text{GL2}\,\lam$ (dashed red) and $2\, E_\text{c}^\text{MP2}\,\lam$ (dashed blue) are plotted for comparison. In the right column, the range extends to $\lam$ large enough for both AC integrands to converge to their asymptotic values, $W_\infty^\ks$ (dashed red) and $W_\infty^\hf$ (dashed blue). The values of $U/|\Delta v|$ selected are 0.1, 1, and 10 (as per the plots' labels). For $U=0.1$ and $|\Delta v|=0.1$ both AC integrands show an almost perfectly linear behaviour within the $\lam$ range from 0 to 1 [panel (a)]. At large $\lam$, their behaviour does not differ very much quantitatively and looks similar qualitatively [(b)]. In the other panels with $U/|\Delta v|=1$ [(e) and (i)] $W_\lam^\ks$ and $W_\lam^\hf$ differ visibly in the range $0\leq\lam\leq 1$. Moreover, in (i-j) there is a striking qualitative difference: the MP adiabatic connection integrand changes curvature twice. The change from concave to convex of $W_\lam^\hf$ is clearly visible in panel (d), where the ratio $U/|\Delta v|=0.1$ and the system is ``weakly interacting''. Note that  $W_\lam^\hf$ starts always above $W_\lam^\ks$. See main text for more discussion. %Finally,  and, in the cases plotted, ends below $W_\lam^\ks$, but the two curves cross beyond $\lam=1$. 
  }
 }
  \end{figure*}
  In Figure~\ref{fig:Wlambda}, we report the two AC integrands, Eqs.~\eqref{eq:wlHD} and~\eqref{eq:wlDFTHD}, in the $0<\lam<1$ range (first column) and for $\lam$ large enough for both AC integrands to converge to their asymptotic values (second column), for selected values of $U$ and $|\Delta v|$ parameters. For $U=1$, we take $|\Delta v| =10, 1, 0.1$, while for $U=0.1$ and $U=10$ we consider only $|\Delta v| =U$. 
  
  The ratio between $U$ and $|\Delta v|$ is an important factor in determining the amount and the type of correlation present in a given calculation.  A dominant $U$ will favor localization of each ``particle'' on a single site (strong interaction or small $|\Delta n|$), whereas a dominant $|\Delta v|$ will favor both particles on one site (weak interaction or large $|\Delta n|$). Nevertheless, the absolute magnitude of $U$ is also important. For example, in the first panel where $U/|\Delta v| =1$ and $U=0.1$, both AC curves show an almost perfect linear behavior in the relevant range between $0$ and $1$. However, for the same ratio and $U=1$ and $10$ [panels (e) and (i)], the AC integrands deviate significantly from linearity (increasingly for larger $ U $).
 An ``almost linear'' AC integrand is considered as an example of \emph{dynamical} correlation whereas a markedly non-linear one as \emph{static} correlation (see e.g. Refs.~\onlinecite{VucFabGorBur-JCTC-20} and~\onlinecite{HelTea-INC-2022}).
  This difference is often assessed globally by looking at the expansion of the correlation energy at small $\lam$, 
  \begin{equation}\label{eq:2nd_ord}
  \lim_{\lam \to 0} E_{c, \lam}^\text{SD} = \lam^2 E_{c, (2)}^\text{SD} + o (\lam^2).
  \end{equation}
In particular, using only the leading order coefficient, $E_{c, (2)}^\text{SD} $, to approximate the correlation energy corresponds to approximating the integrand $ W_\lam^\text{SD} $ as a linear function with slope $2*E_{c, (2)}^\text{SD} $. This coefficient is typically referred to as MP2 and GL2 correlation energies, for the MP and the DFT adiabatic connection, respectively (see Eqs.~\eqref{eq:MPseries} and~\eqref{eq:GLseries}). MP2 and GL2 correlation energies can be used as approximations of the total correlation energy and their performance can be measured through their relative error:
\begin{equation}\label{eq:relerr}
\text{rel. \!err.\! SD} =\Big|\frac{E_{c, (2)} ^\text{SD}-E_\text{c}^\text{SD}}{E_\text{c}^\text{SD}} \Big|\times 100.
\end{equation}
From Figure~\ref{fig:Wlambda}, panel (g), we see that a dominant $U$ does not imply that the AC integrand curve is far from linear or that the relative error is big. In this case, one may talk about ``strong dynamical correlation,'' \changes{in the sense that the correlation energy, though large, is well approximated by a linear AC integrand.} Indeed, the relative error for $U=1$ and $|\Delta v|=0.1$ is 6.0\% and 6.1\% for the MP and the DFT adiabatic connection, respectively. \changes{We can compare this case with the case reported in panel (e) of Figure~\ref{fig:Wlambda}, which has the same value of $U$ and $|\Delta v|=1$. To support the interpretation of the plots in Figure~\ref{fig:Wlambda} and provide more quantitative data, we report in} Table~\ref{tab:1} the values of interacting and HF site-occupation difference, correlation energy, $E_{c, (2)}^\text{SD} /E_c^\text{SD} $ ratio, and relative error, for $U/|\Delta v |= 0.1, \,1,\, 10$ and $U=0.1, \,1,\, 10$. [In the last line of Table~\ref{tab:1}, we also report the quantity $(1-\lam_\text{ext}^\text{SD})$, which will be the focus of section~\ref{sec:accpred}.]

\changes{In the two cases mentioned, which we label (e) and (g) with reference to Figure~\ref{fig:Wlambda}, we see from Table~\ref{tab:1} that the HF correlation energy, $|E_c^\text{HF}|$, is smaller for $|\Delta v|=1$ [(e)] than for $|\Delta v| =0.1$ [(g)]. This is in line with a $U$ dominant over $|\Delta v|$ in the latter case. Likewise, the site-occupation difference is larger for case (e) than for case (g), again in line with a $U$ dominant over $|\Delta v|$ in the latter case, which makes the impact of repulsion stronger there. However, the corresponding relative error, rel. err. HF, in the less correlated case, (e), is more than double that of the more correlated case, (g). A similar trend can be observed when comparing $|E_c^\text{KS}|$ and rel. err. KS in the two cases, with the relative error of the less correlated case, (e), being almost four times that of case (g) in the KS framework.}

Qualitatively, we observe a convex DFT adiabatic connection integrand across the full parameter space of the Hubbard dimer. This result adds to the long list of highly accurate numerical evidence of the (piecewise) convexity of this curve,\cite{ColSav-JCP-99, SavColPoll-IJQC-03, WuYan-JCP-03, TeaCorHel-JCP-09, TeaCorHel-JCP-10, StrKumCorSagTeaHel-JCP-11, VucIroSavTeaGor-JCTC-16, VucIroWagTeaGor-PCCP-17, VucFabGorBur-JCTC-20, HelTea-INC-2022} something which has yet to be proven.  On the contrary, for the MP adiabatic connection, cases have been reported where the integrand is rather concave at small $\lam$.\cite{Per-JCP-18, VucFabGorBur-JCTC-20,DaaGroVucMusKooSeiGieGor-JCP-20} 
However, in the Hubbard dimer setting, we find that the MP adiabatic connection integrand %[eq~\eqref{eq:wlHD}] 
may change curvature twice.  In particular, for any $ U > 0 $, we observe a double change of curvature (DCOC) for a continuous range of $ \Delta v $  beyond a critical value, $|\Delta v_U|>0$, which depends on $U$.  

\begin{figure}
\includegraphics[scale=0.5]{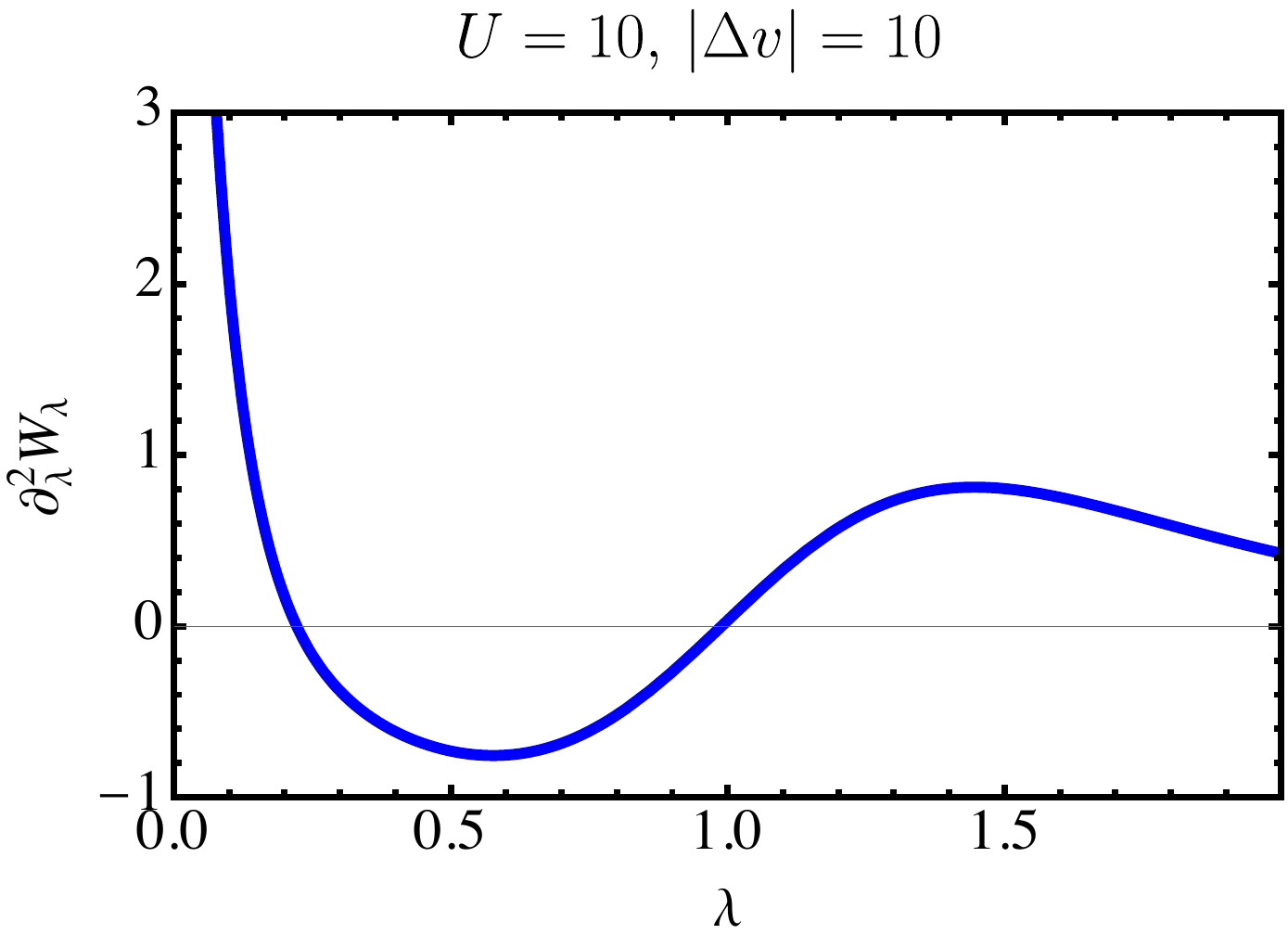}
\caption{\label{fig:dl2WlU10V10}Second derivative with respect to $\lam$ of $W_\lam^\hf$ (also equal to $\frac{\mathrm{d} ^3 E_\lam^\hf}{\mathrm{d} \lam^3}$) for parameters $U=|\Delta v| =10$ [as in panel (i-j) of Figure~\ref{fig:Wlambda}].}
\end{figure}

For example, in panel (i),  
$W_\lam^\hf$ (blue solid curve) starts convex, turning concave around $\lam \approx 0.22$ and turning convex again around $\lam \approx 0.99$. \changes{See also Figure~\ref{fig:dl2WlU10V10}, which shows its second derivative and explicitly demonstrates this DCOC}.

 The change in curvature from concave to convex of $W_\lam^\hf$ is extremely visible in panel (d), where $\Delta v$ is dominating and the system is in the weak-interaction regime.
Note that the integrand of eq~\eqref{eq:wlHD} is convex (i.e. lying \emph{above} its tangent) around $ \lam \to 0 $  for any pair of $ \{ U, \Delta v\} $. In fact, in the cases in which $W_\lam^\hf$ is concave in some region [e.g., for the cases plotted in panels (d) and (i-j)], the MP adiabatic connection integrand still starts convex, having to change curvature an even number of times to reach a bound asymptotic value. 
This is at variance with the mentioned cases where a change of curvature had been previously observed,\cite{Per-JCP-18, VucFabGorBur-JCTC-20,DaaGroVucMusKooSeiGieGor-JCP-20} for which the adiabatic connection integrand is concave (i.e. lying \emph{below} its tangent) around zero (the He atom is one such example\cite{VucFabGorBur-JCTC-20}).

\begin{figure}
\includegraphics[width=\linewidth]{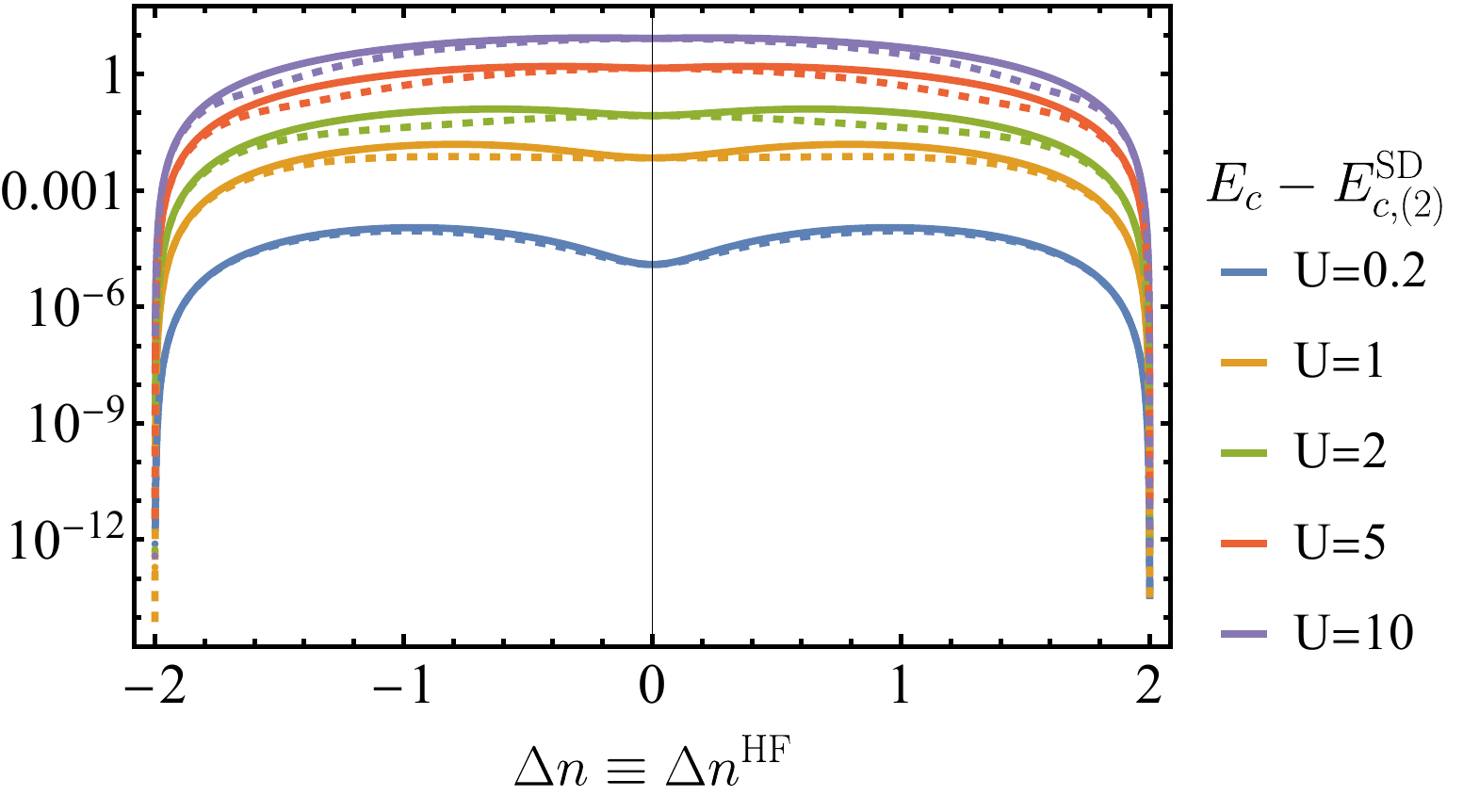}
\caption{Difference between total correlation energy and its second-order expansion, $E_c-E_{c,(2)}^\text{SD}$, with SD = HF (dashed) and KS (solid) for $U=0.2,1,2,5,10$ as functions of the HF and interacting site-occupation difference, respectively. The difference is always positive across all the site-occupation domain, meaning that the second-order expansion always overshoots the correlation energy (logarithmic scale for readability).} 
\label{fig:Ec_Ec2}
\end{figure}

  Oftentimes, whether $ E_c^\text{MP2}$ overestimates or underestimates the exact correlation energy $E_c^\hf $ has been considered~\cite{SeiGiaVucFabGor-JCP-18, VucFabGorBur-JCTC-20, DaaGroVucMusKooSeiGieGor-JCP-20} as an indicator for the convex or concave nature of the curve at the origin; however, this reasoning only holds if the curvature changes at most once in the range $0\leq\lam\leq 1$, something which our findings show to not always hold true. 
Nonetheless, in the Hubbard dimer setting, we find that $ E_c^\text{MP2}$ always overestimates in magnitude the correlation energy, in line with the naive expectation that an AC integrand that is convex at small $\lam$ indicates an $ E_c^\text{MP2}$ that overshoots the correlation energy (see Figure~\ref{fig:Ec_Ec2}).

  In conclusion, 
  in the Hubbard dimer model at any correlation regime we have
\begin{equation}\label{eq:Ec2overshoots}
|E_{c, (2)}^\text{SD}| \geq |E_\text{c}^\text{SD}|
\end{equation}
for ``SD'' either KS (AC integrand convex everywhere) or HF (AC integrand convex for $\lam=0$).
% and the $E_c^\text{MP3}$ ``correction'' is always in the right direction, since $E_c^\text{MP3}\geq0$ as per eq~\eqref{eq:Ecmp3HD}. 

Moreover, as noted in Ref.~\onlinecite{GiaPri-JCP-22}, $E_{c, (2)}^\text{SD}$ has formally the same expression in the two adiabatic connections, namely
    \begin{equation}\label{eq:ecsd2ineq}
    E_{c, (2)}^\text{SD}=-\frac{1}{256} U^2 \left(4-x^2\right)^{5/2}
    \end{equation}
with $x=\Delta n^\text{SD}$. %An inequality between the two second-order approximations can be derived.
 In turn, $|\Delta n^\hf| \geq |\Delta n^\ks|$ and $0\leq |\Delta n^\text{SD}| < 2$. Then, for a given $\{U, \Delta v \}$ pair,
\begin{equation}\label{eq:MP2GL2ineq}
|E_\text{c}^\text{GL2} |\geq |E_\text{c}^\text{MP2} |.
\end{equation}
%In addition, one finds that, if $W_\lam^\hf$ and $W_\lam^\ks$ cross each other, they do so always beyond $\lam =1$, due to eq~\eqref{eq:Ec2overshoots} and to the fact that $|E_\text{c}^\hf| \leq |E_\text{c}^\ks|$ (by definition of $\Phi^\hf$ as the Slater determinant that minimises the total energy).

The thorny question that remains is what causes the observed double change of curvature in $W_\lam^\hf$ and whether it can be expected in Coulomb systems.
As seen, the DCOC in the Hubbard dimer setting occurs in a continuous range of $|\Delta v|$, starting from $|\Delta v_U|$ up to $|\Delta v |\to \infty$. \changes{The value of $|\Delta v_U|$ gets closer to $U$ as $U$ increases. In other words, the DCOC starts appearing around the switch between strong- and weak-interaction regimes and persists in the weak-interaction regime}.
Although we have not found a simple explanation for the DCOC, we highlight in the following some aspects of the MP adiabatic connection that might endow it with such ``extra flexibility'' compared to the DFT one.

Let us start by analyzing the leading coefficient in the weak-interaction expansion of the correlation energy. 

Despite the formal equivalence of $E_{c, (2)}^\text{SD}$in the two adiabatic connections [eq~\eqref{eq:ecsd2ineq}], how this energy term is parsed into the individual components is quite different between the DFT and the MP case.
   To see this, let us introduce the definition of the individual components of the total correlation energy as
\begin{subequations}\label{eq:indcomp}
\begin{align}
U_{c, \lam}^\text{SD}& =  \langle \Psi_\lam^\text{SD} |\lam \, \hat{\mathcal{U}} | \Psi_\lam^\text{SD} \rangle-\langle \Psi_0^\text{SD} |\lam \, \hat{\mathcal{U}} | \Psi_0^\text{SD} \rangle \label{eq:Ucl}\\
T_{c, \lam}^\text{SD} &=  \langle \Psi_\lam^\text{SD} | \hat{\mathcal{T}} | \Psi_\lam^\text{SD}  \rangle-\langle \Psi_0^\text{SD} | \hat{\mathcal{T}} | \Psi_0^\text{SD} \rangle \label{eq:Tcl}
\\
V_{c, \lam}^\text{SD} &=  \langle \Psi_\lam^\text{SD}  | \hat{\mathcal{V}}^{\lam,\text{SD}} | \Psi_\lam^\text{SD}  \rangle-\langle \Psi_0^\text{SD} | \hat{\mathcal{V}}^{\lam,\text{SD}}  | \Psi_0^\text{SD} \rangle  \label{eq:Vcl}\\
 &= \frac{\Delta v^{\lam, \text{SD}}}{2} \left(\Delta n^{\lam,\text{SD}} -\Delta n^\text{SD} \right)\notag
\end{align}
\end{subequations}
With $\hat{\mathcal{V}}^{\lam,\text{SD}} = \sum_i v_i^{\lam,SD} \hat{n}_i$. (Note that, for each individual contribution to correlation, the analogue of eq~\eqref{eq:scalecorr} holds true.)
The $\lam \to 0$ expansion of these terms looks formally identical to eq~\eqref{eq:2nd_ord} and in both HF and KS references, we have
\begin{equation}\label{eq:ec2uc2half}
E_{c, (2)} ^\text{SD} = \frac{U_{c, (2)}^\text{SD}}{2}.
\end{equation}
\changes{Equation~\eqref{eq:ec2uc2half} is a well-known relation of the density-fixed adiabatic connection (i.e., for SD=KS) and has been proven for the Hubbard dimer in this framework in ref~\onlinecite{CarFerSmiBur-JPCM-15} together with the other known relation:}
\begin{equation}\label{eq:tc2uc2half}
T_{c,(2)}^\ks = - \frac{U_{c, (2)}^\ks}{2}.
\end{equation}
Combination of~\eqref{eq:ec2uc2half}  and~\eqref{eq:tc2uc2half} gives $T_{c,(2)}^\ks+ U_{c, (2)}^\ks =E_{c, (2)}^\ks $.  However, in the MP adiabatic connection framework, the non-zero $V_{c, (2)}^\hf$ term complicates the relation between $T_{c,(2)}^\hf$ and $U_{c,(2)}^\hf$ significantly. There, we have
\begin{subequations}
\begin{align}
&T_{c,(2)}^\hf = \frac{\left(4-5 x^2\right) \left(4-x^2\right)^{5/2}}{1024}\label{eq:Tc2}
\\
&V_{c,(2)}^\hf =\frac{5 x^2 \left(4-x^2\right)^{5/2}}{1024},\label{eq:Vc2}
\end{align}
\end{subequations}
such that
\begin{equation}
T_{c,(2)}^\hf + V_{c,(2)}^\hf = - \frac{U_{c, (2)}^\hf}{2}
\end{equation}
\changes{and $T_{c,(2)}^\hf + V_{c,(2)}^\hf + U_{c, (2)}^\hf = E_{c, (2)}^\hf$. Equation~\eqref{eq:ec2uc2half} for SD=HF and expressions~\eqref{eq:Tc2} and~\eqref{eq:Vc2} are obtained by series expanding the terms $U_{c, \lam}^\text{SD}$, $T_{c, \lam}^\text{SD}$, and $V_{c, \lam}^\text{SD}$ [eqs~\eqref{eq:Ucl}-\eqref{eq:Vcl}] around $\lam \to 0$ and retaining only the order $\lam^2$. In fact, in the case of the HF reference, these are known analytically as functions of the density $x=\Delta n^\hf$.}

In conclusion, the presence of the term \changes{$V_{c,(2)}^\hf$ inside $E_{c,(2)}^\hf$} and, in general, of $V_{c, \lam}^\hf$ \changes{inside $E_{c, \lam}^\hf$}, appears to be possibly ``the'' crucial difference between the DFT and the MP adiabatic connections.

Another way to rephrase this crucial difference is that the DFT AC integrand only contains the two-body operator, whereas the MP AC integrand also contains the one-body operator corresponding to the external potential correction [compare eqs~\eqref{eq:wlDFTHD} and~\eqref{eq:wlHD}].
\changes{However, looking at the two contributions in eq~\eqref{eq:wlHD}, $\langle \Psi_\lam^\hf |\mathcal{\hat{U}}|  \Psi_\lam^\hf \rangle$ (two-body) and $\langle \Psi_\lam^\hf |-U\! \sum_i  \frac{n_i^\hf}{2} \hat{n}_i|  \Psi_\lam^\hf \rangle$ (one-body), separately (neglecting the constant term in the equation, as it does not affect the shape of the integrand), provides only limited insight on the occurrence of the DCOC. As a matter of fact, the two-body contribution is everywhere positive and \textit{decreasing} with $\lam$, while the one-body contribution is everywhere negative and \textit{increasing}. Thus, the slope of the former is everywhere negative and that of the latter everywhere positive. In turn, monotonicity of the slope of the total $W_\lam^\hf$ means that the integrand stays convex, and non-monotonicity means that it changes curvature. Unfortunately, predicting whether the total function is monotonic from the slopes of the one-body and two-body contributions is not obvious, since the monotonicity of the resulting curve depends on their relative magnitude along $\lambda$.}
%The two-body contribution (positive) is \textit{decreasing} with $\lam$, while the one-body contribution (negative) is \textit{increasing}. Thus, the slope of the former is everywhere negative and that of the latter everywhere positive.
%In turn, the slope of the total adiabatic connection integrand corresponds to the second derivative of the $\lam$-dependent energy with respect to $\lam$, $\mathrm{d}^2 E_\lam^\hf /\mathrm{d} \lam^2 $. If this derivative is monotonic, the integrand will stay convex, otherwise it will change curvature.
%The delicate interplay between the rate of change of the one-body and two-body contributions determines whether the total function will be monotonic.

%
\begin{table*}[t]
  \centering
  \begin{ruledtabular}
  \renewcommand{\arraystretch}{1.1}
 \begin{tabular}{l|lll|lll|lll}
 & \multicolumn{3}{c}{$U=0.1$}&\multicolumn{3}{c}{$U=1$}&\multicolumn{3}{c}{$U=10$}\\
    $U/|\Delta v|$&0.1&1&10&0.1&1&10&0.1&1&10\\
     \hline
    $|\Delta v|$&1&0.1&0.01&10&1 (e)&0.1 (g)&100&10&1\\
    $\Delta n^\hf $& 1.363 &0.181 & 0.018 & 1.988 & 0.938 & 0.100 &2.000 &1.686& 0.182\\
   $\Delta n$&1.363 &0.180 & 0.018 & 1.988 & 0.775 & 0.068 &2.000 &0.981& 0.004\\
   $|E_\text{c}^\hf|$&2.5 $\times 10^{-4}$ &1.2 $\times 10^{-3} $& 1.3 $\times 10^{-3}$& 1.8 $\times 10^{-6} $& 0.060 & 0.117 &1.9$ \times 10^{-9}$ &0.305& 4.054\\
   $|E_\text{c}^\ks|$&2.5 $\times 10^{-4}$ &1.2 $\times 10^{-3} $& 1.3 $\times 10^{-3}$& 1.8 $\times 10^{-6} $& 0.068 & 0.117 &1.9$ \times 10^{-9}$ &1.145& 4.098\\
   $E_\text{c}^\text{MP2}/E_\text{c}^\hf$&1.033&1.001& 1.001 & 1.109 & 1.125 & 1.060 &1.111 &1.843& 3.020\\
   $E_\text{c}^\text{GL2}/E_\text{c}^\ks$&1.035&1.001& 1.001 & 1.109 & 1.230 & 1.061 &1.111 &5.483& 3.050\\
    rel.\! err.\! HF&3.3\%&0.1\%& 0.1\% & 10.9\% & 12.5\% & 6.0\% &11.1\%&84.3\%& 202\%\\
   rel.\! err.\! KS&3.5\%&0.1\%& 0.1\% & 10.9\% & 23.0\% & 6.1\% &11.1\%&448.3\%& 205\%\\
  % $\lam_\text{ext}^\hf$&0.953&0.998& 0.999 & 0.858 & 0.856 & 0.894 &0.855 &0.591& 0.198\\
   % $\lam_\text{ext}^\ks$&0.950&0.997& 0.999 & 0.857 & 0.711 & 0.892 &0.855 &0.102& 0.196\\
   $1-\lam_\text{ext}^\hf$&0.047&0.002& 0.001 & 0.142 & 0.144 & 0.106 &0.145 &0.409& 0.802\\
   $1-\lam_\text{ext}^\ks$&0.050&0.003& 0.001 & 0.143 & 0.289 & 0.108 &0.145 &0.898& 0.804\\
  \end{tabular}
  \end{ruledtabular}
  %External potential difference, 
  \caption{HF and KS site occupation differences, $\Delta n^\hf$ and $\Delta n$, correlation energies, $E_\text{c}^\hf$ and $E_\text{c}^\ks$, ratio between second-order and exact correlation energy, $E_\text{c}^\text{MP2}/E_\text{c}^\hf$ and $E_\text{c}^\text{GL2}/E_\text{c}^\ks$, relative error on correlation energy, ``rel.\! err.\! SD'' [eq~\eqref{eq:relerr}] and $( 1-\lam_\text{ext}^\text{SD} )$ [see eq~\eqref{eq:lext}], for $U=0.1,\, 1,\,10$ and $U/|\Delta v|=0.1,\, 1,\,10$. This choice corresponds to moving from ``weak-'' to ``strong-interaction'' regime for a fixed $U$, as can be appreciated from the decreasing values of $\Delta n$ and $\Delta n^\hf$ and the increasing values of $E_\text{c}^\hf$ and $E_\text{c}^\ks$. The labels (e) and (g) visible in the third line of the table ($|\Delta v|$) have been added to help readability when discussing these cases in the main text and they refer to the panels of Figure~\ref{fig:Wlambda} with the same labels.}
  \label{tab:1}
\end{table*}

\subsection{Analysis of correlation indicators and accuracy predictors}\label{sec:accpred}
\begin{figure*}
\centering
 \begin{tabular}[c]{cc}
	 {\begin{subfigure}{0.5\textwidth}
      \includegraphics[scale=0.58]{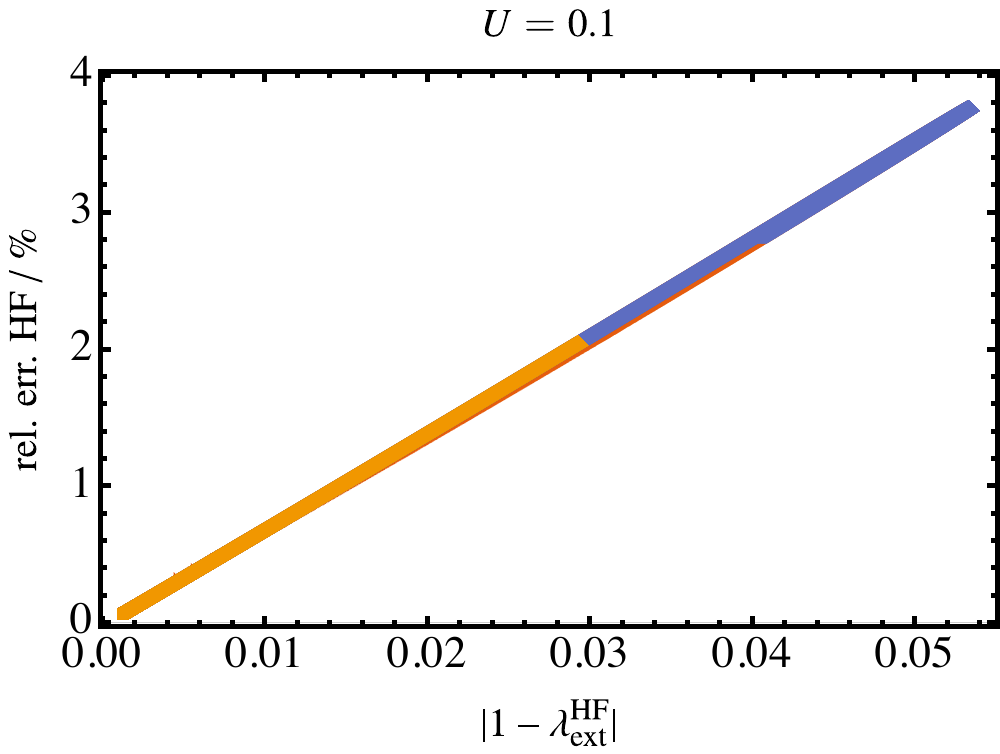}
    \end{subfigure}      \vspace{0.2cm}
} & {\begin{subfigure}{0.5\textwidth}
 \includegraphics[scale=0.58]{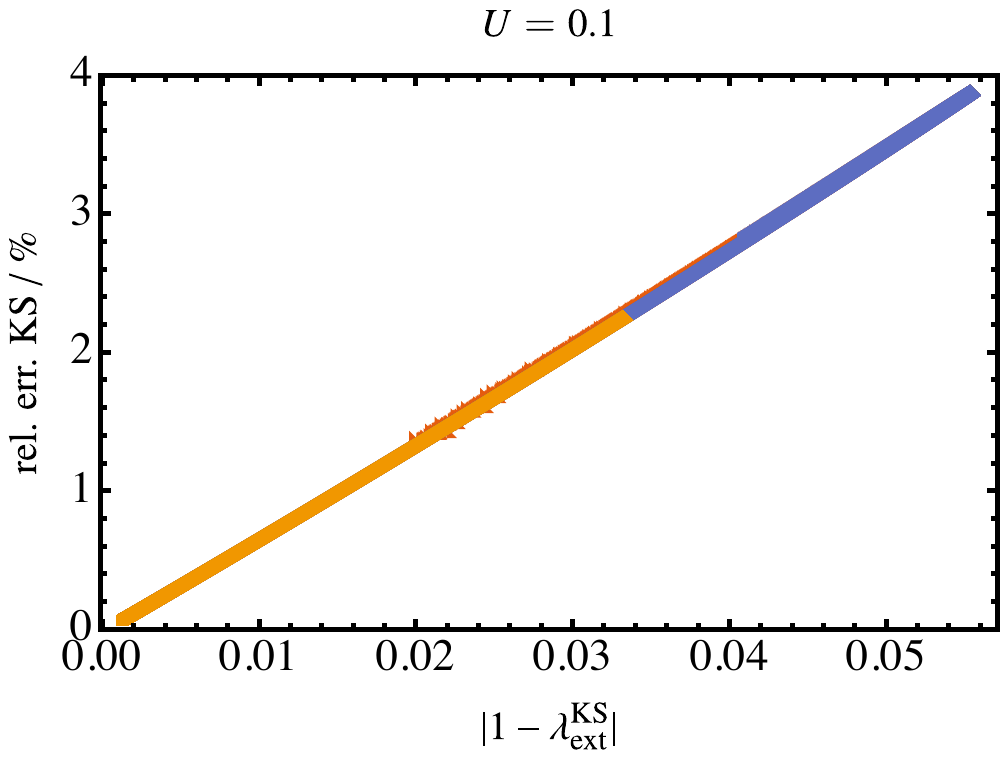}
 \end{subfigure}      \vspace{0.2cm} }\\
  {\begin{subfigure}{0.5\textwidth}
 \phantom{OO} \includegraphics[scale=0.6]{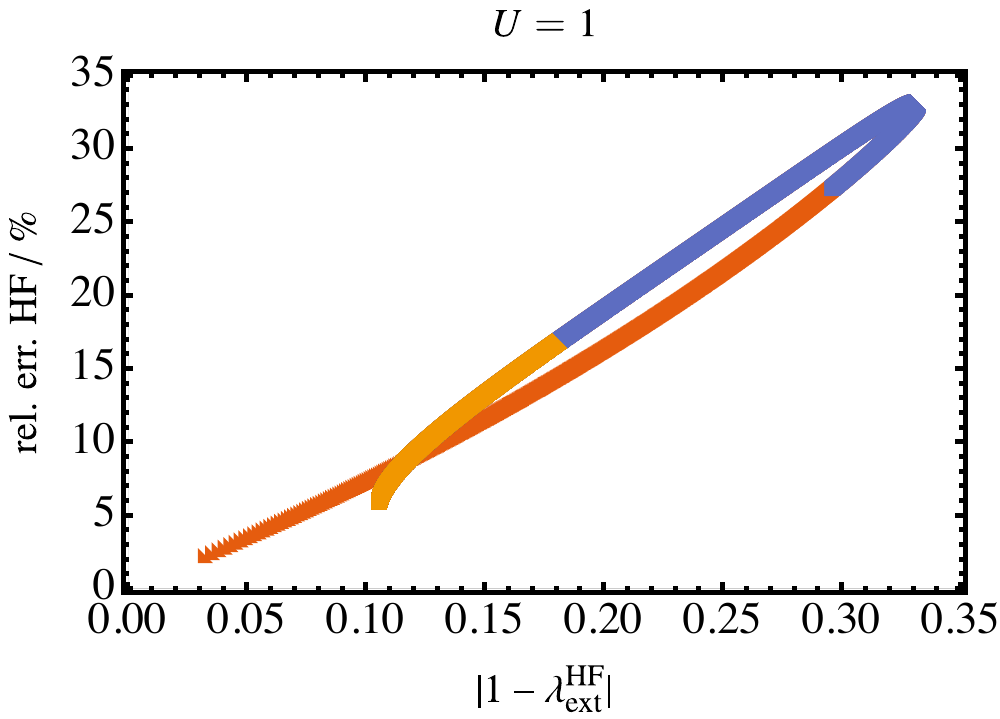}%
 \end{subfigure}\phantom{oo}      \vspace{0.2cm} }
 & {\begin{subfigure}{0.5\textwidth}
 \includegraphics[scale=0.6]{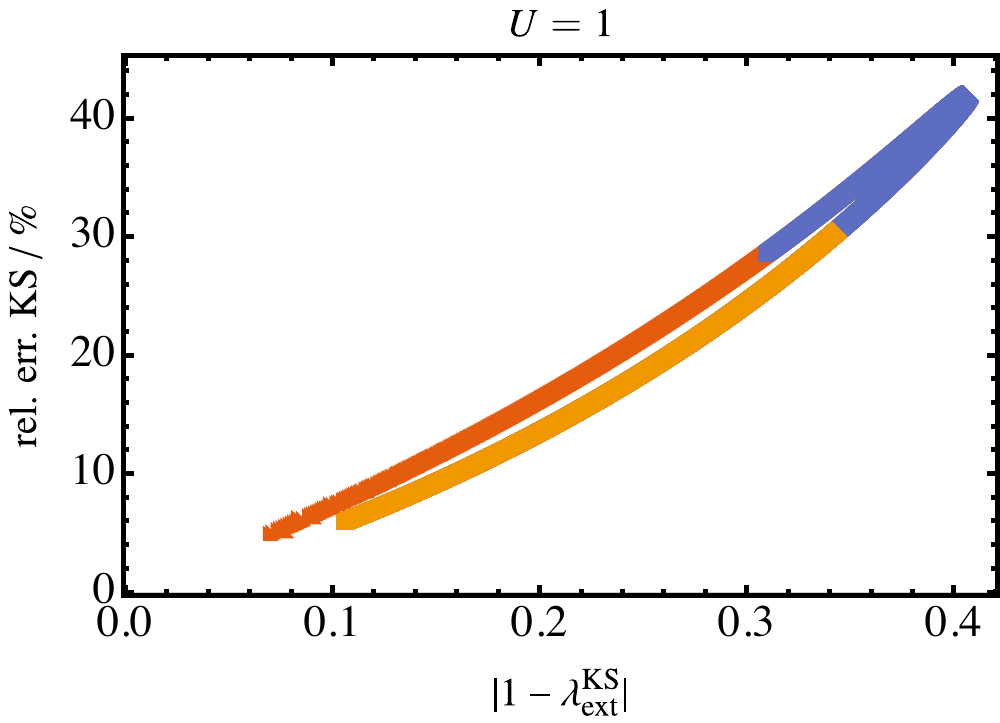}%
 \end{subfigure}       \vspace{0.2cm} }\\
 {\begin{subfigure}{0.5\textwidth}
      \includegraphics[scale=0.6]{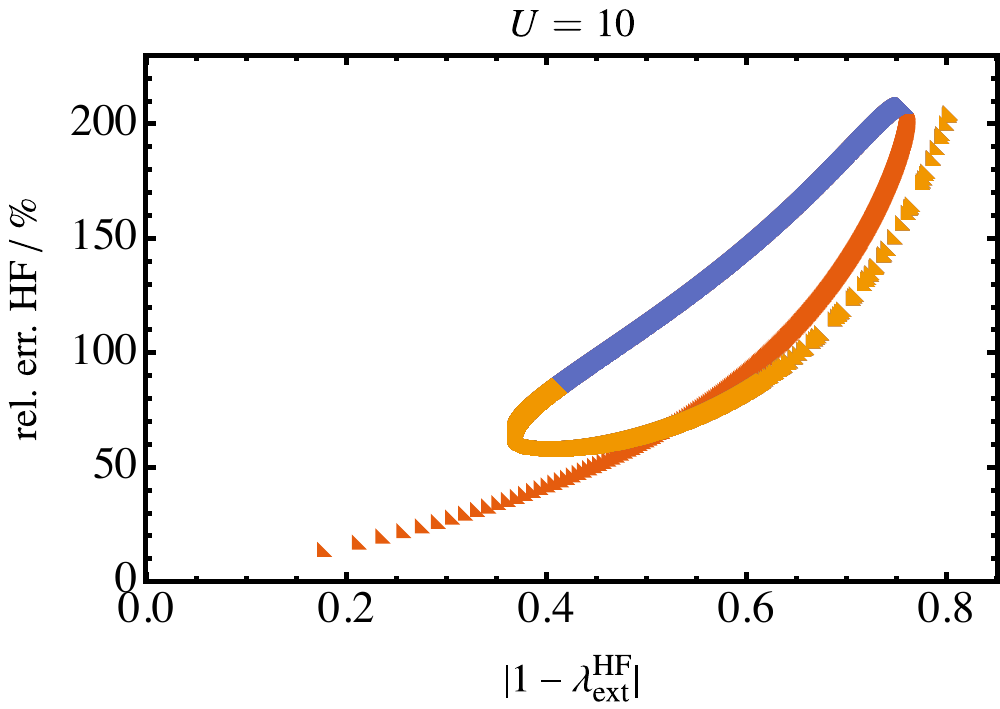}
    \end{subfigure}
} & {\begin{subfigure}{0.5\textwidth}
 \includegraphics[scale=0.6]{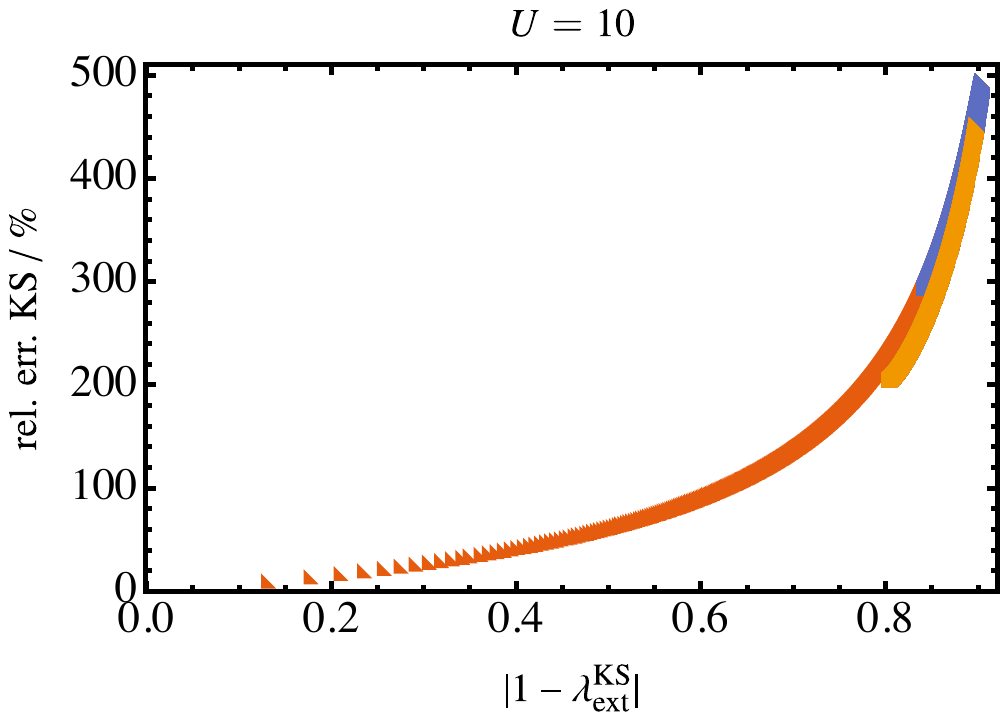}
 \end{subfigure}}
 \end{tabular}   
 \caption{\label{fig:relerrvsAP} %\small
 \footnotesize{Relative error in the second-order estimate of the correlation energy \textit{vs} $|1-\lam_\text{ext}|$ for different values of on-site interaction strength $U=0.1, 1, 10$ and full domain of interacting site-occupation difference: $0<\Delta n<2$ (data points' spacing is $0.0001$). The full domain is split into subdomains: $0 \leq \Delta n\leq 1 $ (yellow), $1 \leq \Delta n\leq 1.9 $ (blue), and $1.9 \leq \Delta n\leq 2 $ (red). Left column concerns the M\o ller-Plesset adiabatic connection and $E_\text{c}^\hf$, while right column, the DFT adiabatic connection and $E_\text{c}^\ks$. 
 }
 }
  \end{figure*}
% all other things equal
In this section, we calculate the correlation indicator $\lam_\text{ext}$,~\cite{VucIroWagTeaGor-PCCP-17,VucFabGorBur-JCTC-20} used in the context of adiabatic connection methods, in the full parameter space of our model for both adiabatic connections.
This indicator is defined as
\begin{equation}\label{eq:lext}
\lam_\text{ext}^\text{SD}:=\frac{W_1^\text{SD}}{2\,E_{c, (2)} ^\text{SD}}.
\end{equation}
This is a dimensionless quantity that determines the $\lam$ value at which the linear curve, given by the slope of $W_\lam^\text{SD}$ at $\lam =0$, crosses the constant curve corresponding to $W_{c,1}^\text{SD}$, where $W_{c,1}^\text{SD} =W_{1}^\text{SD} - W_{0}^\text{SD}$.  %In other words, this is the value of the AC integrand at $\lam =1$ minus its value at $\lam=0$. 
(However, in our model, $W_{0}^\text{SD}=0$, therefore $W_{c,1}^\text{SD} =W_{1}^\text{SD} $.)

By definition, $\lam_\text{ext}^\text{SD}=1$ means that the curve $2\,E_{c, (2)}^\text{SD} \lam$ crosses the constant $W_1^\text{SD}$ exactly at $\lam =1$. 
In the assumption that the MP AC integrand changes curvature at most once in the relevant range of $\lam$ between zero and one, $\lam_\text{ext} ^\text{SD}= 1$ is indeed enough to say that the second-order energy expansion recovers all the correlation energy and that the AC integrand is linear and fully ``dynamically correlated.'' Thus, how much $\lam_\text{ext}^\text{SD}$ differs from one, i.e., the quantity $|1-\lam_\text{ext}^\text{SD}|$, has been considered as a predictor of the accuracy of the second-order perturbation estimate of the correlation energy. This can be measured using the relative error defined in eq~\eqref{eq:relerr}.
However, the possible presence of more than one change of curvature invalidates this deduction. Our results therefore raise the interesting question of whether the quantity $|1-\lam_\text{ext}^\hf|$ remains a meaningful predictor of the corresponding relative error.

In addition to this possible issue arising from the double change of curvature, the correspondence between a certain non-zero value of $|1-\lam_\text{ext}^\text{SD}|$ and the relative error is not obvious. Even with the assumption that there is at most \emph{one} change of curvature in the relevant range of $\lam$ between 0 and 1 holding (and therefore $\lam_\text{ext}^\text{SD} = 1$ implies that the relative error is exactly zero), 
 %, in other words, the answer to the question``what value of relative error corresponds to a $|1-\lam_\text{ext}|=0.05$?'' is unclear. Therefore, 
 it is useful to extend our analysis to the DFT adiabatic connection as well, which appears to always satisfy such assumption, to further clarify this correspondence and its applicability.

%makes no difference between W_{c,1} and W_1 cause W_0=0 for us anyways - maybe you should specify your use of hartree and exchange in thr intro - so taking them out of the SIL section
%and the accuracy predictor derived from it - $|1-\lam_\text{ext}|$ -

A technical aspect worth mentioning is that the accuracy predictor $|1-\lam_\text{ext}|$, introduced in reference~\onlinecite{VucFabGorBur-JCTC-20}, has been applied in the context of the MP adiabatic connection in terms of a modified correlation indicator, $\lam_\text{ext}^\text{SPL}$.  ``SPL'' refers to the Seidl-Perdew-Levy formula,~\cite{SeiPerLev-PRA-99} which approximates the DFT AC integrand using a convex interpolant and thus serves as a means to approximate $W_{c,1}$ (which is typically unknown in its exact form). This interpolation formula has proven surprisingly useful for molecular applications when used in the framework of the MP AC.~\cite{FabGorSeiDel-JCTC-16, VucGorDelFab-JPCL-18, GiaGorDelFab-JCP-18, VucFabGorBur-JCTC-20, DaaFabDelGorVuc-JPCL-21}
In the above applications, however, the SPL interpolation formula is used as a correction not to absolute energies, but rather to energy differences, as its accuracy has been found to be poor for the former but satisfactory for the latter. Thus, also the use of the accuracy predictor, if one uses the SPL or similar formulas to approximate $W_{c,1}$, is recommended in terms of energy differences and not absolute energies.
In the present work, however, $W_{c,1}$ (or $W_1$) can be calculated \emph{exactly} and therefore can be used directly to construct $|1-\lam_\text{ext}|$ for absolute energies, in line with the original idea \cite{VucIroWagTeaGor-PCCP-17} that the relative error [eq~\eqref{eq:relerr}] increases as $\lam_\text{ext}$ deviates from one.
Finally note that, by virtue of eq~\eqref{eq:Ec2overshoots}, $\lam_\text{ext} \leq 1$ and the modulus can be ignored in the context of this work. 

% Its interpretation as ``accuracy predictor'' follows the idea that the relative error [eq~\eqref{eq:relerr}] increases the more $\lam_\text{ext}$ deviates from one, as originally proposed \cite{VucIroWagTeaGor-PCCP-17}. %following the original idea \cite{VucIroWagTeaGor-PCCP-17} that the relative error [eq~\eqref{eq:relerr}] increases the more $\lam_\text{ext}$ deviates from one.

%cit the interaction energy HF adiabatic connection curve is probably, most of the times, convex and, in general, well-modeled by the difference between two convex curves
%  
In Figure~\ref{fig:relerrvsAP}, we plot the relative error on the y-axis and $|1-\lam_\text{ext}|$ on the x-axis for the same $\{ U, \Delta n\}$ pair, exploring $U=0.1,\,1,\,10$  and the full site occupation difference range, $0\leq\Delta n\leq 2$, within the MP (left column) and the DFT (right) adiabatic connections.
From the plots, we see that the relative error is not a function of the quantity $|1-\lam_\text{ext}|$; however, this quantity does appear to be a significant factor in determining the relative error.
Moreover, although there is a somewhat commensurate behavior between the two columns, there are both quantitative and qualitative differences between the results for the MP and DFT adiabatic connections. Quantitatively, the error given by GL2 is both more sensitive to $U$ and typically larger than the error given by MP2. The same is true if we look at $|1-\lam_\text{ext}^\ks|$ and $|1-\lam_\text{ext}^\hf|$. This trend can be more easily appreciated by looking at Table~\ref{tab:relerr+AP} where the mean, $\mu$, of the distribution of the relative error as well as of the $|1-\lam_\text{ext}|$ for each given $U$ over the full $\Delta n$ range is calculated, together with their standard deviations, $\sigma$. In particular, we see that $\mu_\text{rel.\!err.KS}> \mu_\text{rel.\!err.HF}$ and $\mu_{|1-\lam_\text{ext}^\ks|}> \mu_{|1-\lam_\text{ext}^\hf|}$ in all cases, and we see that the increment of the KS quantities with $U$ is much larger than that of the HF quantities. %It is also interesting to notice that the standard deviation of the distributions is consistently of the same order of magnitude as their mean.

The qualitative differences between the two columns of Figure~\ref{fig:relerrvsAP} emerge by looking at how smaller $\Delta n$ values are placed with respect to the relative error for the cases where fixed $|1-\lam_\text{ext}|$ yields two branches. For example, in the first two rows ($U=0.1,\,1$), we see that, for a fixed $\lam_\text{ext}$, a lower $\Delta n$ (typically) corresponds to a \emph{larger} relative error for HF (left column) and to a \emph{smaller} relative error for KS (right column). Since a smaller $\Delta n$ may be associated with stronger interaction, it seems that ``more weakly-interacting cases'', meaning $1.9\leq \Delta n \leq 2$ (in red), may be somehow more problematic than more strongly-interacting, meaning $0 \leq \Delta n \leq 1$, in the DFT adiabatic connection. %, which is somewhat unexpected.
In the case of the MP adiabatic connection, this swap in the expected order happens only for few values in the first two panels. For example, for $U=1$ and at $|1-\lam_\text{ext}^\hf| \equiv 0.11$, the red curve is above the yellow one.
For $U=10$, the situation becomes even more involved, as
up to three different values of relative error may correspond to a given $|1-\lam_\text{ext}^\hf|$. While for intermediately interacting cases [~$1\leq\Delta n\leq1.9$ (in blue)], the relative error appears to be higher than for weakly interacting cases [~$1.9\leq\Delta n\leq 2$ (in red)], strongly interacting cases [~$0\leq\Delta n\leq1$ (in yellow)], appear to correspond to the lowest relative error.

To conclude, it seems that $|1-\lam_\text{ext}|$ is an important contribution to the relative error even in the presence of a double change of curvature, however not in a straightforward way. The question, ``What relative error, $y$, corresponds to a $|1-\lam_\text{ext}| = x$?'' is ill-posed, in the sense that rather a \emph{range} of relative error, $y\pm \Delta y$, seems to correspond to a value $x$. This range is different depending on whether we are considering the MP or DFT adiabatic connections and also depends on $U$.
Unwrapping the dependence of the relative error on $|1-\lam_\text{ext}|$ may be complicated, but the picture that we get from investigating these quantities in the Hubbard dimer setting seems to recommend caution when using this latter as an accuracy predictor.

On the positive side, it appears that, if $|1-\lam_\text{ext}|<0.1$, the relative error remains below $10\%$ in all cases, regardless of $U$. \changes{(Note that this is in line with the findings of ref~\onlinecite{VucFabGorBur-JCTC-20} regarding the S22 and S66 datasets for non-covalently bonded complexes.)}
Moreover, if a pair of $\{U, \Delta n \}$ makes $|1-\lam_\text{ext}|<0.1$ in one adiabatic connection, the same holds true for the other. This is because the behavior of the two adiabatic connections differs more for larger $|1-\lam_\text{ext}|$. Consequently, $|1-\lam_\text{ext}|<0.1$ is the range in which this predictor is more consistent between the two adiabatic connections.
%A valid question that remains is how much the findings obtained from this model are transferable to actual molecular systems.
%
\begin{table}[b]
\caption{\label{tab:relerr+AP} Mean, $\mu$, and standard deviation, $\sigma$, for the distributions of the relative error and of $|1-\lam_\text{ext}|$ accross the full site-occupation range, $0\leq\Delta n\leq 2$, for $U=0.1,\,1,\,10$ in the MP and DFT adiabatic connections (``HF'' and ``KS'', respectively).}
\begin{ruledtabular}
\begin{tabular*}{0.8\columnwidth}{@{\extracolsep{\fill}}lcccccc}
& \multicolumn{2}{c}{$U=0.1$}&\multicolumn{2}{c}{$U=1$}&\multicolumn{2}{c}{$U=10$}\\
& HF & KS & HF & KS & HF & KS \\
 \hline
$\mu_\text{rel.\! err.}$&1.92&2.04 &  17.84&  25.85 & 105.81& 379.72\\
$\sigma_\text{rel.\! err.}$&2.33 & 2.45 &  20.22 &  28.78 & 115.71 & 392.20\\
$\mu_{|1-\lam_\text{ext} |}$&0.028 & 0.030 &  0.198 &  0.290 & 0.502 & 0.872 \\
$\sigma_{|1-\lam_\text{ext} |}$ & 0.033 & 0.035 &  0.213 &  0.308 & 0.517  & 0.873\\
\end{tabular*}
\end{ruledtabular}
\end{table}
\subsection{The $\lam$-dependent site occupation difference}\label{sec:lSOD}
\begin{figure}
\includegraphics[scale=0.4]{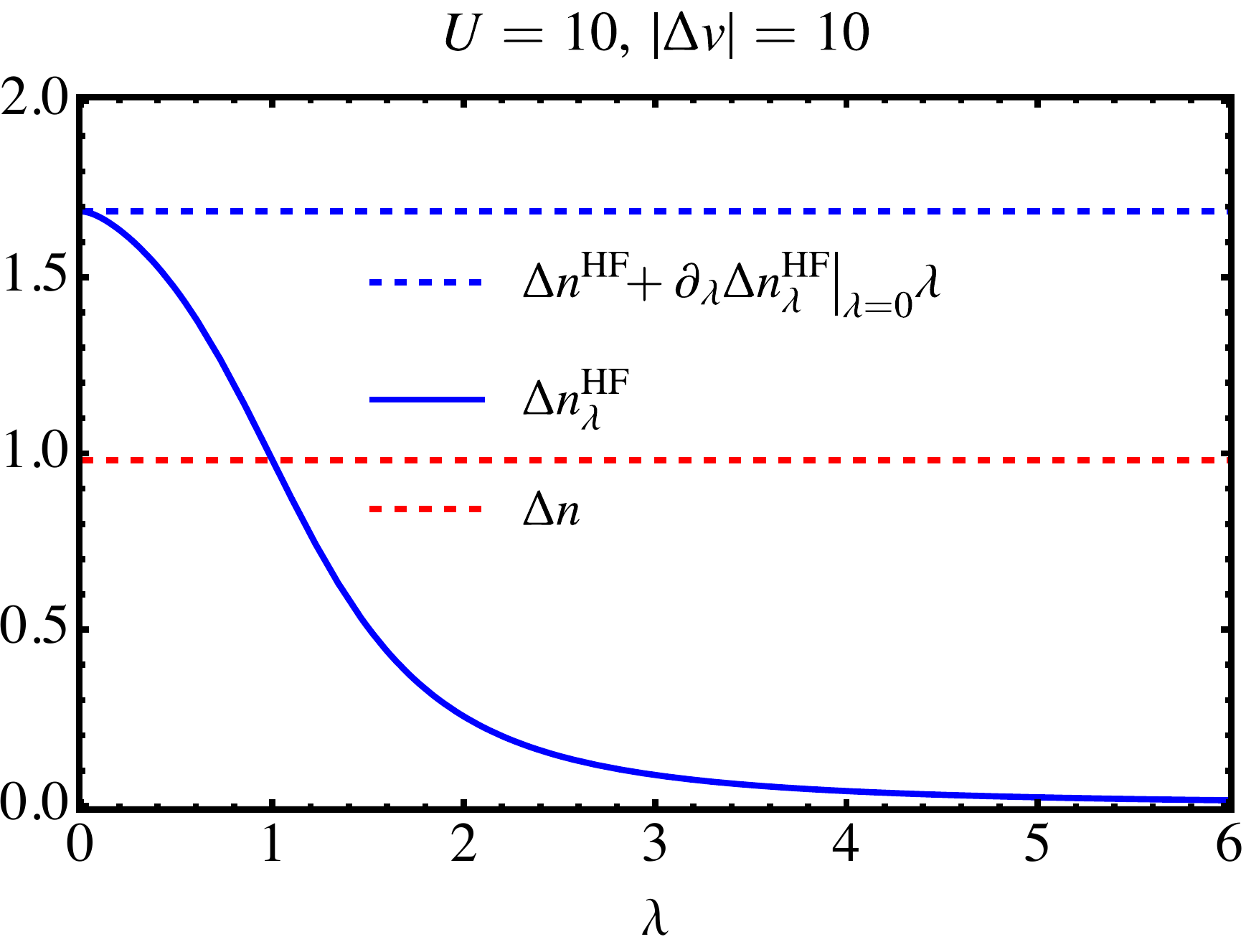}
\caption{Evolution of the $ \lam$-dependent site occupation, $ \Delta n^\hf_\lam $, along $ \lam $ for $ U =10 $ and $|\Delta v | =10 $. The tangent to the curve at $ \lam=0 $ is plotted in dashed blue, while the interacting site occupation, $ \Delta n $, in dashed red.} 
\label{fig:Nlambda}
\end{figure}

In this section, we focus solely on the $\lam$-dependent site occupation difference in the MP adiabatic connection, since the site-occupation difference is kept fixed in the DFT adiabatic connection by construction.
Quite conveniently, in the Hubbard dimer, this quantity can be expressed analytically at any $\lam$ in the full $\{U, \Delta v\}$ parameter space (although the expression is lengthy and we eschew reporting it here).
The evolution of $\Delta n^\hf_\lam$ along $ \lam$ is shown in Figure~\ref{fig:Nlambda}.
Even though the plot only shows $ \Delta n^\hf_\lam $ for $U=|\Delta v |=10$, the properties that can be observed in the figure are not specific to these values. The first thing we notice is that 
\begin{equation}
\lim_{\lam \to \infty} \Delta n^\hf_\lam \sim 0 \quad \forall \,\, U, \, \Delta v <\infty.
\end{equation}
Namely, for any finite $ \Delta v $ (or any  $ \Delta n^\hf < 2$ ), the effect of the repulsion enhanced by $ \lam $ is asymptotically dominating, confining one particle on each site.
Secondly, we see that the first-order derivative of the site occupation in $ \lam $ is zero at $ \lam=0 $, 
\begin{equation}\label{eq:BTV}
\frac{\partial \Delta n^\hf_\lam}{\partial \lam}\Big|_{\lam = 0} = 0, 
\end{equation}
meaning that the HF site occupation is stable under first-order variations of the coupling parameter. 
To see why this is the case, we use the expression of the interacting wavefunction according to perturbation theory up to first order,
\begin{equation}\label{eq:Psil_1stOrd}
|\Psi_\lam^\hf \rangle \sim |\Psi_0^\hf \rangle + \lam \sum_{i \neq 0} \frac{\langle \Psi_{0,i}^\hf | \hat{\mathsf{V}} |  \Psi_0^\hf\rangle}{E_{0}^\hf -E_{0,i}^\hf} |\Psi_{0,i}^\hf\rangle,
\end{equation}
where $\hat{\mathsf{V}} $ is the perturbation operator, corresponding to $ U\,  \sum_i \left( \hat{n}_{i\uparrow} \hat{n}_{i\downarrow} -\frac{n_i^\hf}{2}\hat{n}_i\right) $ in the Hubbard dimer [compare eq~\eqref{eq:ACHFham}], and where the extra subscript $i$ in $\Psi_{0,i}^\hf$ and $E_{0,i}^\hf $ indicates the spectrum of excited states and corresponding energies of the HF reference system. We omit it when indicating the GS and its energy (in other words, we write $\Psi_0^\hf$ instead of $\Psi_{0,0}^\hf$). 
In the Hubbard dimer, the summation is finite and exhausted with only two terms: $i=1$ and $i=2$, corresponding to the first and second excited states. Then, the slope of the site occupation difference around $\lam=0$ is given by:
\begin{eqnarray}\label{eq:nl_1stOrd}
\frac{\partial \Delta n^\hf_\lam}{\partial \lam}\Big|_{\lam = 0} & = & 2 \left(\frac{\langle \Psi_{0,1}^\hf | \hat{\mathsf{V}} |  \Psi_0^\hf\rangle}{E_{0}^\hf -E_{0,1}^\hf}\langle \Psi_0^\hf |\reallywidehat{\Delta n} | \Psi_{0,1}^\hf \rangle + \right.\notag \\
& &  \left. \frac{\langle \Psi_{0,2}^\hf | \hat{\mathsf{V}} |  \Psi_0^\hf\rangle}{E_{0}^\hf -E_{0,2}^\hf}\langle\Psi_0^\hf |\reallywidehat{\Delta n} | \Psi_{0,2}^\hf \rangle \right)
\end{eqnarray}
where $\reallywidehat{\Delta n}=\hat{n}_1 - \hat{n}_0$.
The first term in the summation is zero by virtue of Brillouin's theorem which makes the numerator $\langle \Psi_{0,1}^\hf | \hat{\mathsf{V}} |  \Psi_0^\hf\rangle$ vanish, as the first excited state corresponds to a singly excited determinant. On the contrary, the term $\langle \Psi_{0,2}^\hf | \hat{\mathsf{V}} |  \Psi_0^\hf\rangle$ is non-zero, but the expectation value $\langle\Psi_0^\hf |\reallywidehat{\Delta n} | \Psi_{0,2}^\hf \rangle $ is, owing to the occupation of the doubly excited state being exactly reversed compared to the ground state. %(the doubly excited HF state can be regarded as a dark state).

\changes{Equation~\eqref{eq:Psil_1stOrd} is not specific to the Hubbard dimer, but is valid for different systems depending on the expression of the perturbation operator $\hat{\mathsf{V}}$. For electronic structures in real space, for instance, we know this to be $\hat{\mathsf{V}} =\hat{V}_{ee} - \hat{V}_\text{HF}$ [eq~\eqref{eq:MPac}].
As for eq~\eqref{eq:nl_1stOrd}, one must first consider that many more terms would appear in the summation on the right hand side (infinitely many, in theory, and a number depending on the basis-set size, in practice). However, given that and}
translating now the site-occupation-difference operator, $\reallywidehat{\Delta n}$, into the density operator in real space, $\sum_i^N \delta \left( \br -\br_i \right)$, one may then expect that the $\lam$-dependent density will be flat around $\lam=0$ also for two-electron systems in real space whenever the transition probability density $\langle\Phi^\hf |\sum_i^N \delta \left( \br -\br_i \right) | \left( \Phi^\hf\right)_{ij}^{ab} \rangle $ between any doubly-excited HF configuration and the ground HF state is negligible (or exactly zero).
 
\section{Strong-interaction limits and inequalities}\label{sec:SIL}

For the Hubbard dimer, the asymptotic Hamiltonian $ \hat{H}_\infty^\hf $ introduced in eq~\eqref{eq:PsiHFinfdef} reads
\begin{equation}\label{eq:HinfMP}
\mathcal{\hat{H}}_\infty^\hf =U \sum_i \left( \hat{n}_{i \uparrow}\hat{n}_{i \downarrow} -\frac{n_i^\hf}{2}\hat{n}_i\right)
\end{equation}
which is diagonal in the adopted basis, the asymptotic eigenstates being simply $(100)^T,(010)^T,(001)^T$. 
Let us now compare the MP asymptotic Hamiltonian with the DFT one.

Although we do not know the expression of the $ \lam$-dependent external potential $\Delta v^{\lam, \ks} $ at each $ \lam $ in closed form, we know the large-$ \lam $ behaviour of $F_\lam$ to be~\cite{CarFerSmiBur-JPCM-15}
\begin{equation}
F_\lam (\Delta n) \sim \lam \,\frac{U}{2}|\Delta n| \quad \quad \quad \lam \to \infty.
\end{equation}

Then $\lim_{\lam \to \infty} \frac{\Delta v^{\lam, \ks}}{\lam} = \Delta v^\infty = - U\, \text{sgn}\left(\Delta n \right) $, and the asymptotic Hamiltonian introduced in eq~\eqref{eq:DFTwfinf} reads in this case
\begin{eqnarray}\label{eq:Hinfdft}
\mathcal{\hat{H}}_\infty^\ks & = & U \sum_i  \hat{n}_{i \uparrow}\hat{n}_{i \downarrow}- \frac{U}{2}\text{sgn}\left(\Delta n\right)\left(\hat{n}_1-\hat{n}_0 \right).
\end{eqnarray}
Just as $ \mathcal{\hat{H}}_\infty^\hf $, $ \mathcal{\hat{H}}_\infty^\ks $ is diagonal in the adopted basis.

In Table \ref{tab:EinftyX}, we report the expectation value of the asymptotic Hamiltonian, $\hat{H}_\infty^\text{SD}$ for the two ACs in each of the basis vectors.
\begin{table}[b]
\caption{\label{tab:EinftyX} Expectation value $\langle \hat{H}^\text{SD}_\infty\rangle$ evaluated on each of the basis vectors for SD = HF, KS.}
\begin{ruledtabular}
\renewcommand{\arraystretch}{1.3}
\begin{tabular*}{0.8\columnwidth}{@{\extracolsep{\fill}}lcc}
& MP & DFT  \\
 \hline
$(100)^T$ &$\frac{U}{2}  \Delta n^\hf $  &  $ U \left(1+\text{sgn}(\Delta n) \right) $\\
$(010)^T$ & $ -\frac{U}{2}\Delta n^\hf $  &$ U \left(1-\text{sgn}(\Delta n) \right)$\\
$(001)^T $ &$ -U $  &  $ 0  $\\
\end{tabular*}
\end{ruledtabular}
\end{table}
For the MP AC, the GS corresponds to the state where each particle is confined on each site, except for $ \Delta n^\hf = \pm 2$, when the state with both particles on the site with lower external potential also contributes. So, for $|\Delta n^\hf | < 2$, the asymptotic wave function corresponds simply to
$ \Psi^\hf_\infty = \left( 0 0 1 \right)^T $ yielding $ \Delta n^\hf_\infty =0 $ as already seen in Figure~\ref{fig:Nlambda}.
On the contrary, for the DFT AC, the ground state is two-fold degenerate, except for $ \Delta n= 0$, for which the sign of the site occupation difference is undefined and both states with two particles on one site contribute. To satisfy the density constraint, we need to make a linear combination which mixes the state with $ \Delta n =0$ with the other relevant state according to the sign of $ \Delta n $. Considering, e.g., only the branch with $ \Delta n >0 $, we have
\begin{equation}\label{eq:PsiSCEHD}
\Psi_\infty^\ks =   \begin{pmatrix} 0 \\
k \\ \sqrt{1-k^2}
\end{pmatrix},
\end{equation}
where $k = \sqrt{\frac{\Delta n}{2}}$. % while the density constraint is satisfied also for k with opposite sign, the minus combination gives higher energy - coming from the hopping term!
This picture is quite different than the usual SCE picture (in real space) where the co-motion functions [eq~\eqref{eq:PsiSCE}] enforce the density constraint. Here, this task is taken over by the coefficients of the basis vectors, which determine how the degenerate ground states are linearly combined.

Let us now consider the asymptotic AC integrands, $ W_\infty^\text{SD} $. 
It is quite instructive to look at how the argument used in reference~\onlinecite{SeiGiaVucFabGor-JCP-18} to prove eq~\eqref{eq:ineqWinf} can be easily adapted to the Hubbard dimer case.
%maybe the general lattice setting, Nofsites=Nofparticles>2, could bring further insight...
We introduce first the bifunctional $ \mathcal{W} (\Delta n, \Delta v) $ as
{\small{\begin{equation}\label{eq:bifctl}
\mathcal{W}(\Delta n, \Delta v)  := \min_\Psi \langle \Psi | \mathcal{\hat{U}} -\frac{\Delta v}{2} \left(\hat{n}_1 -\hat{n}_0 \right)|\Psi \rangle + \frac{\Delta v}{2} \Delta n
\end{equation}
}}

as well as the following definitions
\begin{eqnarray}
& &U_\text{H} (\Delta n) := \frac{U}{2} \left(1+\left( \frac{\Delta n }{2}\right)^2 \right) \label{eq:Hartreelattice} \\
& &\Delta v_\text{H}  (\Delta n) := 2\,\frac{\mathrm{d} \,U_\text{H} (\Delta n)}{\mathrm{d} \Delta n}.  \label{eq:VHlattice}
\end{eqnarray}
%Here we differentiate 
Note that we define the Hartree energy, $U_\text{H}$, and potential, $\Delta v_\text{H}$ as in Ref.~\onlinecite{GiaPri-JCP-22}, although different definitions are possible.~\cite{CarFerSmiBur-JPCM-15}

Using the Legendre-Fenchel transform formulation of SOFT, one finds
\begin{eqnarray}
\mathcal{W}(\Delta n, \Delta v^\infty(\Delta n)) & = &\max_{\Delta v}  \mathcal{W}(\Delta n, \Delta v)  \\
& = & \min_{\Psi\to \Delta n} \langle \Psi | \mathcal{\hat{U}} | \Psi \rangle.\label{eq:VeesceHD}
\end{eqnarray}

On the other hand, plugging the asymptotic wave function $ \Psi_\infty^\ks $ into eq~\eqref{eq:wlDFTHD}, we have
\begin{eqnarray}\label{eq:Winftyks}
W_\infty^\ks(\Delta n) & = & \langle \Psi_\infty^\ks |\,\mathcal{\hat{U}\,}|\Psi_\infty^\ks \rangle -U_\text{H}(\Delta n).
\end{eqnarray}
Since the minimizer in eq~\eqref{eq:VeesceHD} is precisely $ \Psi_\infty^\ks $, we have
\begin{equation}\label{eq:bifctlDFT}
W_\infty^\ks\left(\Delta n \right) + U_\text{H}\left(\Delta n \right)=\max_{\Delta v}  \mathcal{W}(\Delta n, \Delta v).
\end{equation}

In the MP strong-interaction case, we cannot make use of any convex analysis tool. However, substituting the asymptotic wave function $ \Psi_\infty^\hf $ into definition~\eqref{eq:wlHD} for general $ \lambda $ and using Eqs.~\eqref{eq:bifctl},~\eqref{eq:Hartreelattice} and~\eqref{eq:VHlattice}, one obtains
\begin{equation}\label{eq:bifctlMP}
W_\infty^\hf \!\left(\Delta n^\hf \right) + U_\text{H} \!\left(\Delta n^\hf \right) = \mathcal{W}(\Delta n^\hf, \Delta v_\text{H}(\Delta n^\hf) )
\end{equation}
Choosing now $\Delta n \equiv \Delta n^\hf$ and comparing Eqs.~\eqref{eq:bifctlDFT} and~\eqref{eq:bifctlMP} leads to
\begin{equation}\label{eq:ineqinfty}
W_\infty^\hf(\Delta n^\hf) \leq W_\infty^\ks (\Delta n)\Big|_{\Delta n \equiv \Delta n^\hf}
\end{equation}
which is the Hubbard dimer analogue of eq~\eqref{eq:ineqWinf}.

In practice, we can also work out the explicit expressions, which read
\begin{equation}\label{eq:WinfMP}
 W_\infty^\hf(\Delta n^\hf) = \frac{U}{2} \left( \left( \frac{\Delta n^\hf}{2}\right)^2 -1 \right),
\end{equation}
and
\begin{eqnarray}\label{eq:Winftyksexp}
W_\infty^\ks(\Delta n) = -\frac{U}{2}\left( 1-\Big| \frac{\Delta n}{2}\Big|\right)^2
\end{eqnarray}
This latter expression is in agreement with eq~(56) of reference~\onlinecite{CarFerSmiBur-JPCM-15} (as in the \textit{corrigendum} \cite{CarFerSmiBur-JPCM-16}) for $E_c^\ks$ in the $U\to \infty$ limit.
This limit corresponds to the situation where the indirect interaction energy (the on-site repulsion minus the mean field term) becomes dominant and, indeed, we have
$
 E_c^\ks \sim U_c^\ks \sim W_\infty^\ks$ as $ U\to \infty$, while $ T_c^\ks $ is subleading.
In the HF case, we do have
\begin{equation}
 E_c^\hf \sim  W_\infty^\hf \quad \quad \text{for} \quad U\to \infty,
\end{equation}
with $T_c^\hf$ similarly subleading; however in this case the extra term coming from the external potential remains non-negligible at large $U$ and contributes to $ W_\infty^\hf$. 
Introducing the leading-order terms in the large-$\lam$ expansions of the individual components [eq~\eqref{eq:indcomp}],
\begin{subequations}
\begin{align}
\lim_{\lam \to \infty} U_{c, \lam}^\hf& = U_{c, \infty}^\hf \,  \lam + o(\lam)  \\
\lim_{\lam \to \infty} V_{c, \lam}^\hf &=V_{c, \infty}^\hf \, \lam + o(\lam),  
\end{align}
\end{subequations}
 where $U_{c, \infty}^\hf = - U_\text{H}(\Delta n^\hf)$ and $V_{c, \infty}^\hf = - U $, then
\begin{equation}
 W_\infty^\hf =  U_{c, \infty}^\hf + V_{c, \infty}^\hf.
\end{equation}
 \begin{figure}
\includegraphics[width=\linewidth]{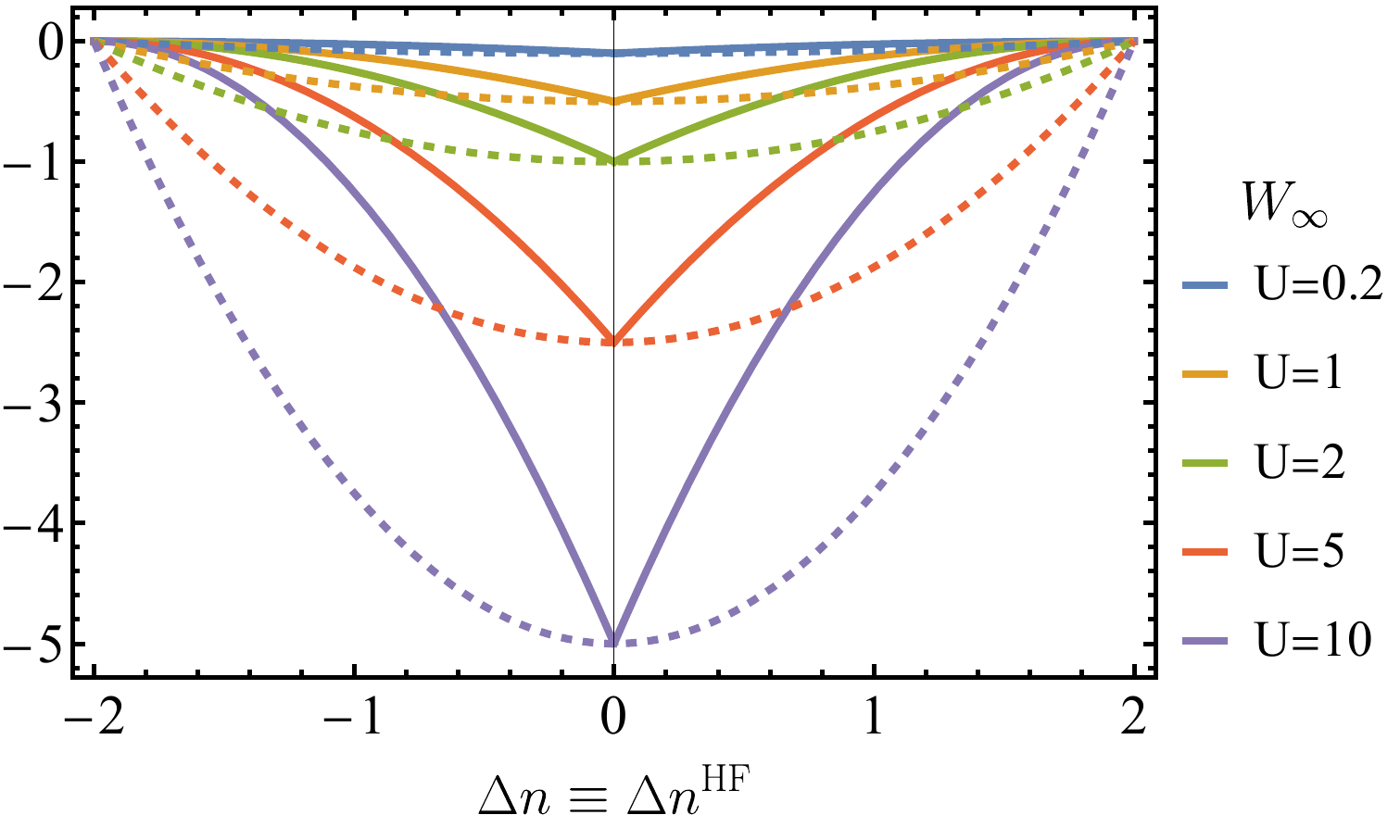}
\caption{Strong-interaction limit as a function of the occupation, $ W_\infty \left( \Delta n \right) $, of the MP (dashed) and  of the DFT (solid) adiabatic connections.} 
\label{fig:Winfvsn}
\end{figure}
The quantities $ W_\infty^\ks$ and $ W_\infty^\hf$ are plotted in fig~\ref{fig:Winfvsv} for the same site-occupation difference; from which the inequality~\eqref{eq:ineqinfty} nicely stands out (the dashed line, which corresponds to $ W_\infty^\hf $, is always below the thick one except at $\Delta n = 0$).
However, considering that both the interacting and the HF site-occupation differences are known analytically as functions of the external potential difference, $\Delta v$, it becomes interesting to look at how the two asymptotic AC integrands relate to one another for the same interacting Hamiltonian (same $\{U, \Delta v\}$ pair), plotted in fig~\ref{fig:Winfvsv}. From the figure, we can see that there is a significant range of $\Delta v$, at fixed $U$, for which the DFT asymptotic AC integrand is actually lower than the MP one.
This may come as a surprise, since, in fig~\ref{fig:Wlambda}, no such cases are shown, but this is only because we chose to plot examples from the three different regimes $U/|\Delta v| =0.1,1,10$. However, choosing, e.g., $U=10$ and $\Delta v = 5$, the DFT asymptotic AC integrand is lower than the MP (compare fig~\ref{fig:Winfvsv}). Indeed, in fig~\ref{fig:exampleWinf} where both AC integrands are plotted along $\lam$ for these parameters, we observe that the MP AC integrand remains always \emph{above} the DFT one.

\begin{figure}
\includegraphics[width=\linewidth]{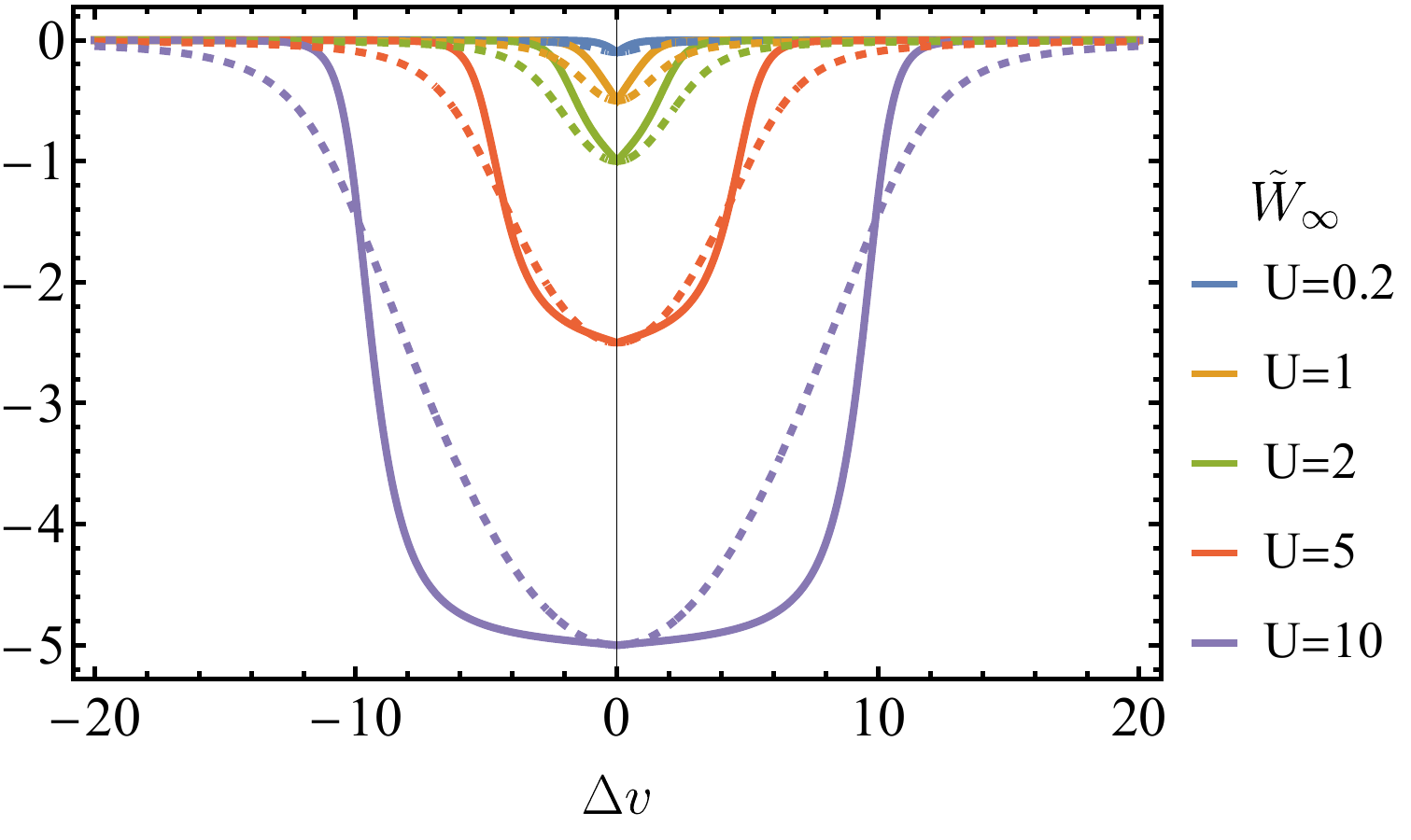}
\caption{Strong-interaction limit as a function of the external potential, $ \tilde{W}_\infty \left( \Delta v \right) $ of the MP (dashed) and  of the DFT (solid) adiabatic connections.} 
\label{fig:Winfvsv}
\end{figure}
\begin{figure}
\includegraphics[scale=0.4]{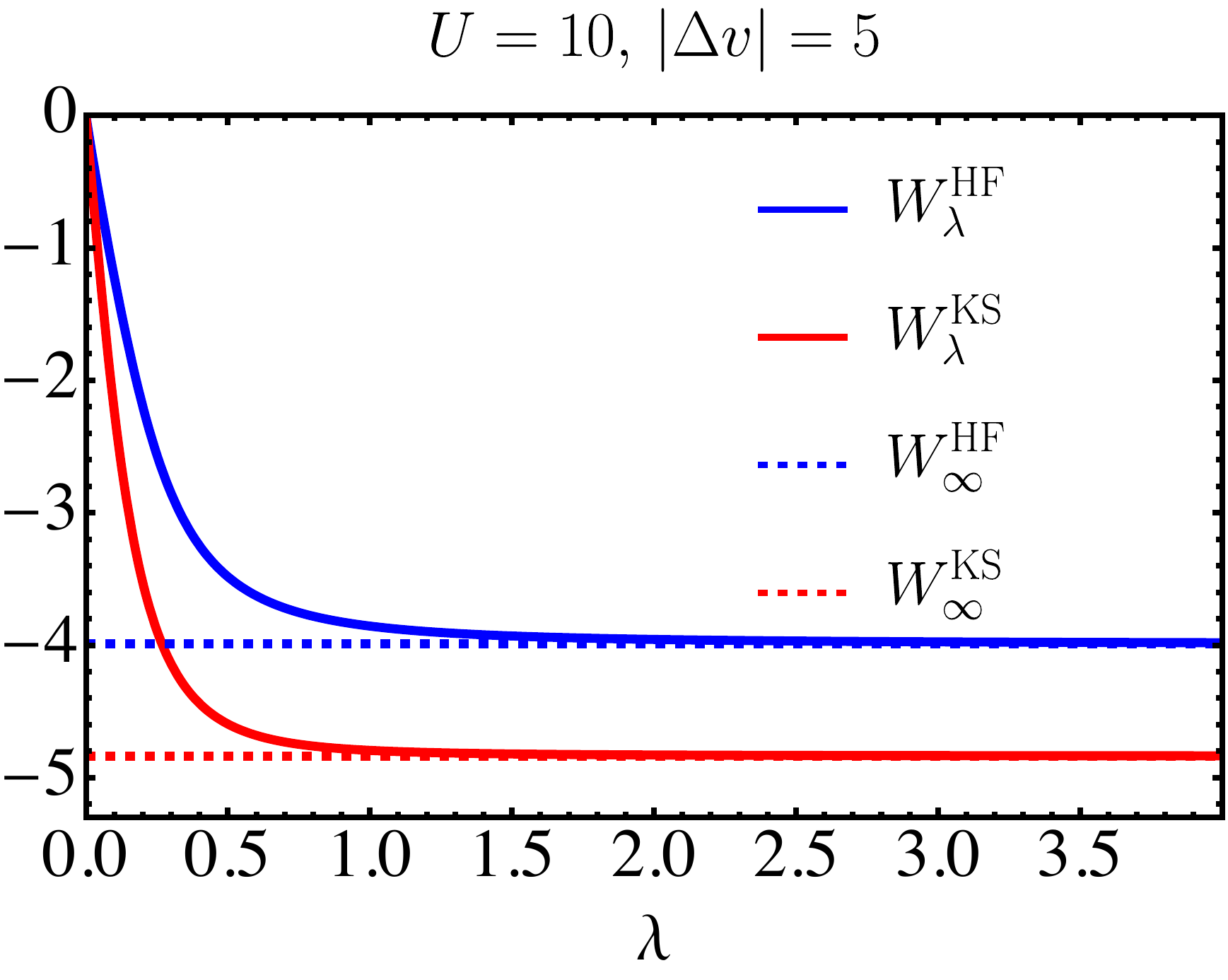}
\caption{Example of $U$ and $\Delta v$ parameters for which $ W_\infty^\ks <  W_\infty^\hf$.} 
\label{fig:exampleWinf}
\end{figure}

\section{Conclusions and perspectives}\label{sec:Conclu}
In this work, we have calculated the M\o ller-Plesset (MP)  and the density-fixed (DFT) adiabatic connection for the asymmetric Hubbard dimer. The Hamiltonian of this model is fully determined by only two parameters, e.g., interaction strength and external potential (or interaction strength and site occupation), allowing one to investigate the two adiabatic connections systematically at different correlation regimes.
The main result we report is that, while the DFT adiabatic connection integrand appears to be convex for any value of the parameters $\{U, \Delta v \}$  (in line with expectations), the MP integrand shows a double change of curvature for a continuous range of $\Delta v$, at any given $U>0$. Since the Hubbard dimer is often considered as a prototype for a stretched diatomic molecule, our finding might signal the presence of such previously unexpected behaviour also in molecular systems. %importance of the novel feature for designing approximations
We have argued that the external potential contribution to $E_c^\hf$ ($V_c^\hf$), which is absent in $E_c^\ks$, may be responsible for the extra flexibility of the MP adiabatic connection over the DFT one, and we have derived an inequality between the MP2 and GL2 correlation energies  [eq~\eqref{eq:MP2GL2ineq}]. We have calculated the accuracy predictor based on $\lam_\text{ext}$ of eq~\eqref{eq:lext} in the complete site-occupation range ($0\leq \Delta n <2$) and contrasted it with the relative error corresponding to the MP2/GL2 correlation energies (Figure~\ref{fig:relerrvsAP}).
 For the MP adiabatic connection,  we have shown that the derivative of the site-occupation with respect to $\lam$ is zero around the HF density, for any $\{ U, \Delta v\}$ pair (section~\ref{sec:lSOD}).
 
Finally, we have characterised the strong-interaction limit in both adiabatic connections for our model. While the asymptotic MP wave function, $\Psi_\infty^\hf$, is simply the symmetry-adapted state with one particle on each site, the DFT asymptotic state must mix in the state with two-particles on the same site as well, with a coefficient determined by the density constraint [eq~\eqref{eq:PsiSCEHD}].
The inequality relating the asymptotic adiabatic connection integrands, $W_\infty^\hf$ and $W_\infty^\ks$, that holds for a given density in real-space [eq~\eqref{eq:ineqWinf}] has been translated in the lattice setting [eq~\eqref{eq:ineqinfty}], and the two quantities have been compared also for a given external potential (see Figure~\ref{fig:Winfvsv}).

As the double change of curvature is an important element to keep in mind in the use of adiabatic connection interpolation methods, a possible next step could be to calculate the MP adiabatic connection for a simple heteronuclear molecule at large internuclear distances and verify whether this feature is present there (other models, such as the Moshinsky atom~\cite{Mos-AJP-68} may be used).
Constructing the MP adiabatic connection integrand for the UHF reference state for the asymmetric Hubbard dimer could also shed some light on the origin of this feature.

Another extension of this work, in the spirit of developing functional approximations that use the HF density as reference, may be to construct a $\lam$-dependent local density approximation from Quantum Monte Carlo data for the uniform electron gas along the MP adiabatic connection. Such data could also reveal whether the double change of curvature and/or the behaviour of the $\lam$-dependent density being flat around the HF density [eq~\eqref{eq:BTV}] are encountered in the uniform electron gas. 

\section{Data availability}
A Mathematica notebook supporting the results presented in this work is available at \url{https://www.hypugaea.com/s/MPDFTAC_SuppNotebook.nb}.

\section{Acknowledgments} Fruitful discussions with Dr. Juri Grossi are gratefully acknowledged. This work is supported by the U.S. Department of Energy, National Nuclear Security Administration, Minority Serving Institution Partnership Program, under Awards DE-NA0003866 and DE-NA0003984. We acknowledge all indigenous peoples local to the site of University of California, Merced, including the Yokuts and Miwuk, and thank them for allowing us to live, work and learn on their traditional homeland (see \href{www.hypugaea.com/acknowledgments}{https://www.hypugaea.com/acknowledgments}).

\appendix
\section{Local performance of Liu-Burke functional}
\begin{figure*}
\centering
 \begin{tabular}[c]{cc}
	 {\begin{subfigure}{0.5\textwidth}
      \includegraphics[scale=0.4]{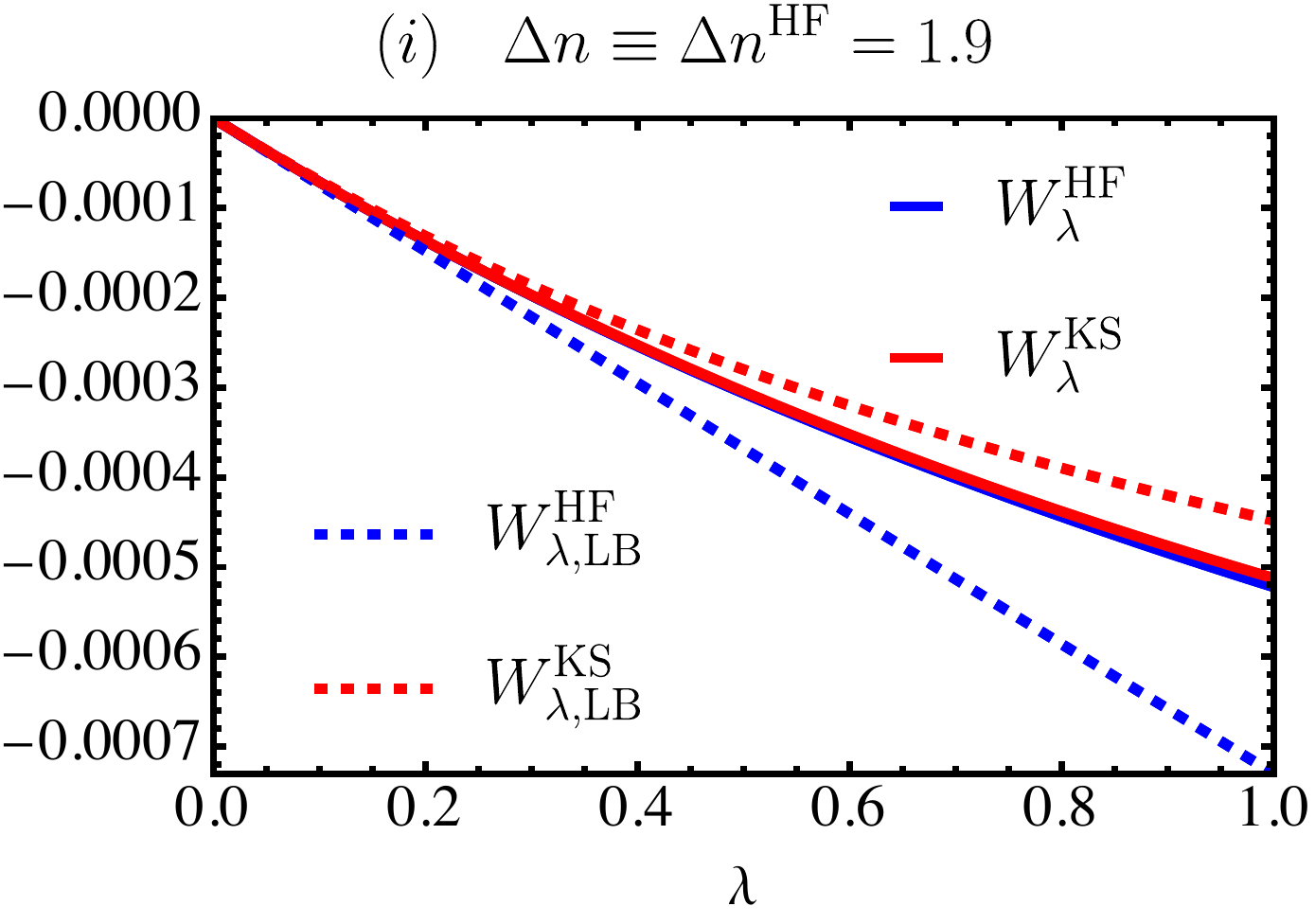}
    \end{subfigure}\phantom{O}
} & {\begin{subfigure}{0.5\textwidth}
 \includegraphics[scale=0.4]{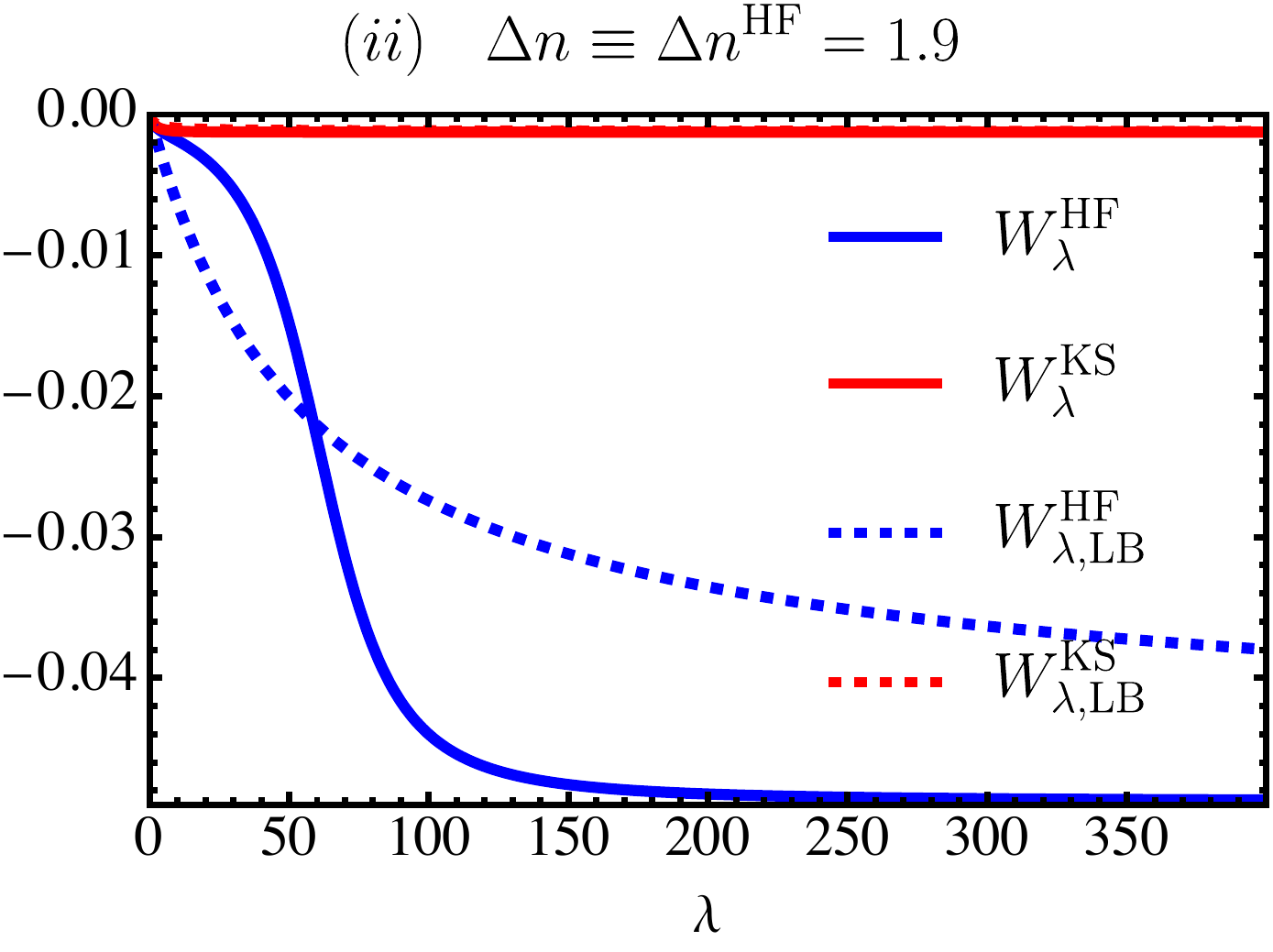}
 \end{subfigure}}\\
  {\begin{subfigure}{0.5\textwidth}
 \includegraphics[scale=0.4]{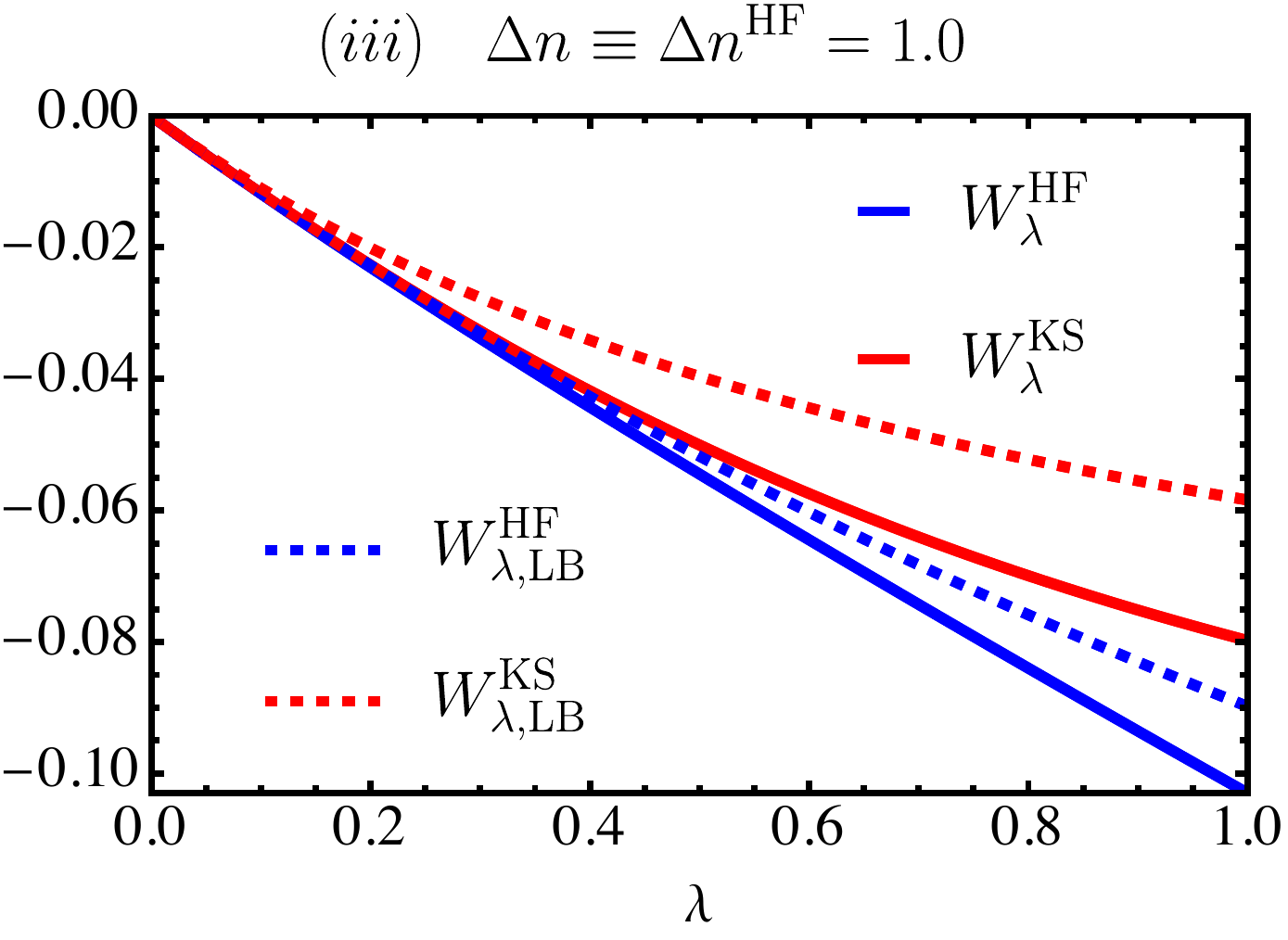}%
 \end{subfigure}\phantom{oo}}
 & {\begin{subfigure}{0.5\textwidth}
 \includegraphics[scale=0.4]{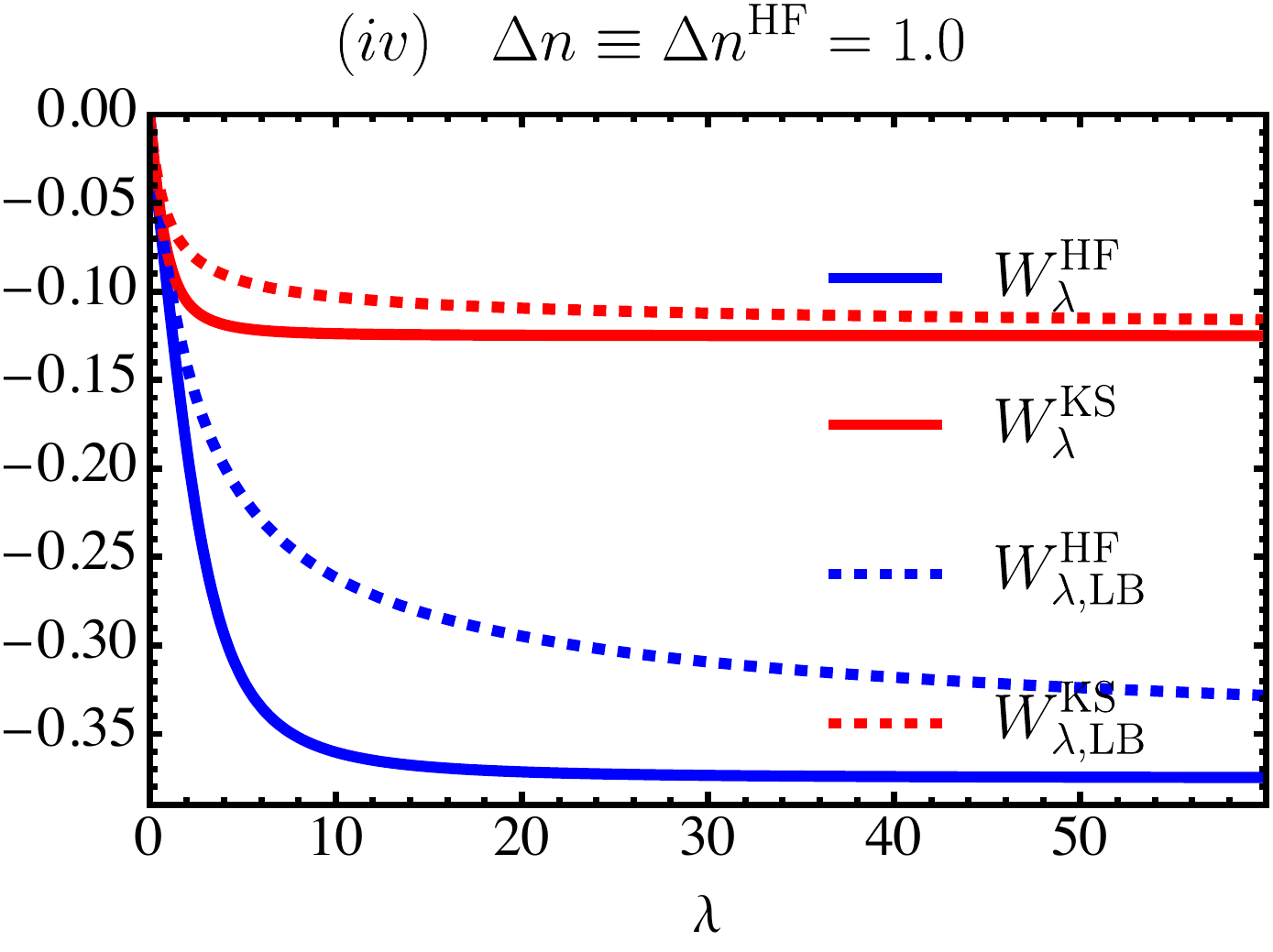}%
 \end{subfigure}}\\
 {\begin{subfigure}{0.5\textwidth}
 \includegraphics[scale=0.4]{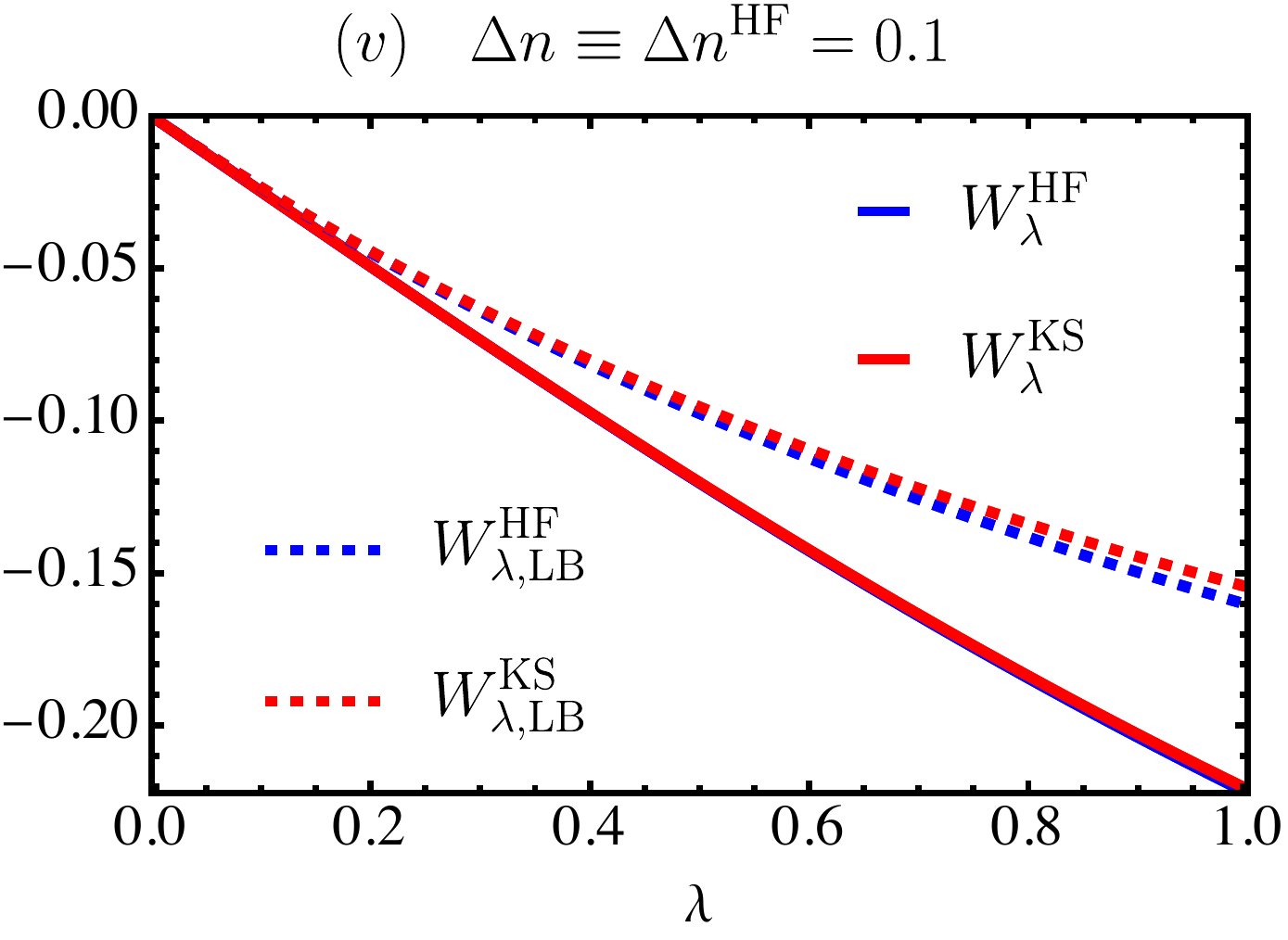}
    \end{subfigure}
} & {\begin{subfigure}{0.5\textwidth}
 \includegraphics[scale=0.4]{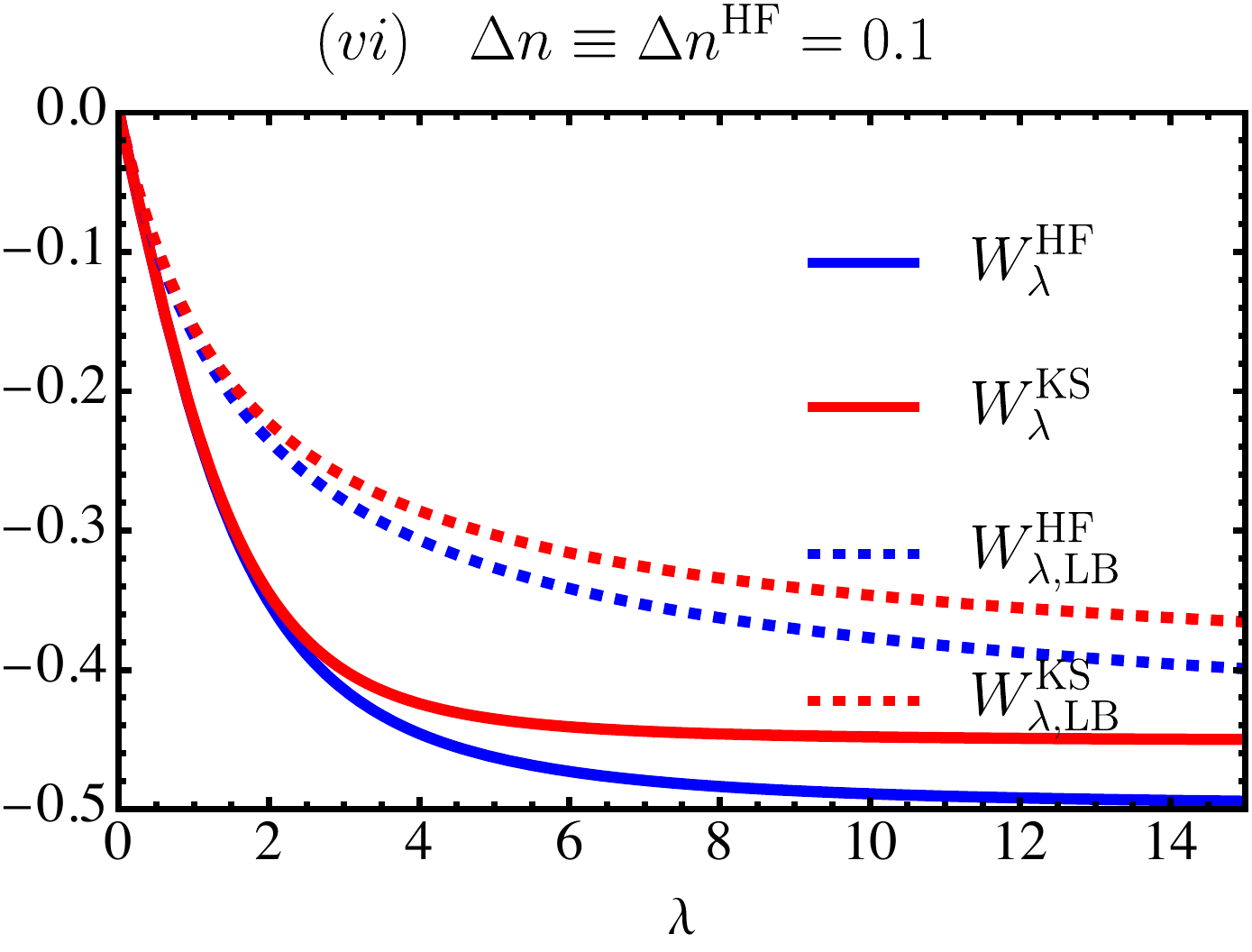}
 \end{subfigure}}
 \end{tabular}   
 \caption{\label{fig:WlambdaLB} %\small
 \footnotesize{Shapes of the two exact AC integrands, $W_\lam^\ks$ (solid red) and $W_\lam^\hf$ (solid blue) in comparison with the LB interpolation formula [eq~\eqref{eq:LB}] (dashed), for $U=1$ and values of the site occupation difference $\Delta n =1.9,\, 1.0,\,0.1$ (from weak- to strong-interaction regime) in the range $0<\lam<1$ (left column) and $\lam>>1$ (right). 
 }
 }
  \end{figure*}

\begin{figure}
\includegraphics[width=\linewidth]{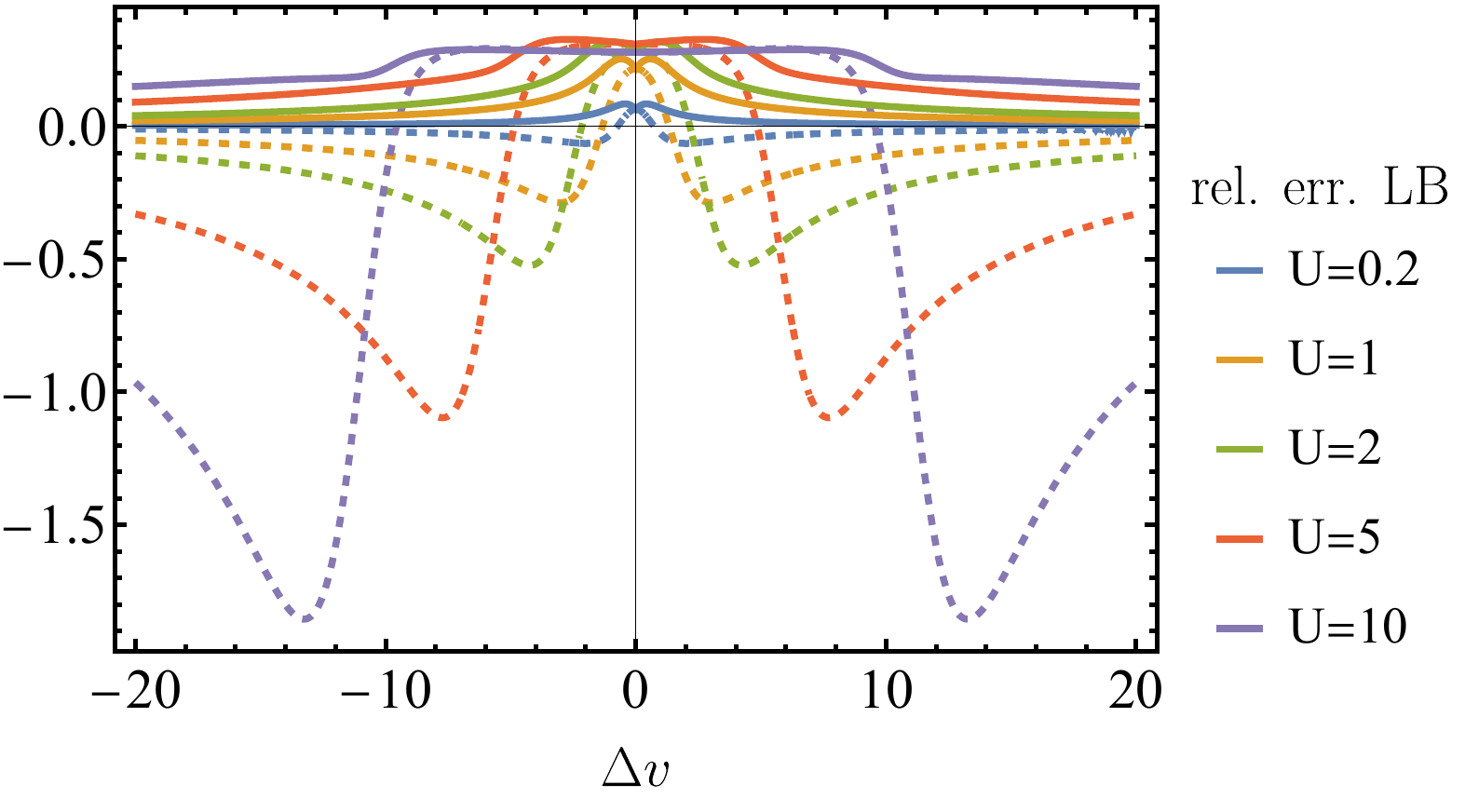}
\caption{Relative error of the LB functional [eq~\eqref{eq:relerrLB}] for the KS (thick) and the HF (dashed) correlation energy. [Compare Figure~11 of ref~\onlinecite{GiaPri-JCP-22}.]} 
\label{fig:relerrLB}
\end{figure}

Although in this work we have calculated the adiabatic connection integrands and other quantities related to them ($E_{c,(2)}^\text{SD}$ and $W_\infty^\text{SD}$) without introducing any approximation, in practical applications these curves are often modeled using interpolation formulas.
In light of this, we briefly investigate the local performance of a representative of this class of density functional approximations in the following.
Consider the Liu-Burke (LB)\cite{LiuBur-PRA-09} formula
{\small{\begin{equation}\label{eq:LB}
  W_\lam^\text{LB}  =-\frac{\tilde{W}_\infty}{2}\left(\frac{1}{\left( 1+\frac{4 W_0'\, \lam}{5 \tilde{W}_\infty}\right)^2} +\frac{1}{\left( 1+\frac{4 W_0'\, \lam}{5 \tilde{W}_\infty}\right)^\frac{1}{2}} \right)+\tilde{W}_\infty,
\end{equation}}}
where $W_0'=2\,E_{c, (2)}^\text{SD}$ and $\tilde{W}_\infty=W_\infty^\text{SD}-E_x^\text{SD}$, except that in the Hubbard dimer $E_x$ can be set to zero and $\tilde{W}_\infty=W_\infty^\text{SD}$.~\cite{GiaPri-JCP-22}
Because the weak-interaction ingredient, $E_{c,(2)}^\text{SD}$, is formally the same in both MP or DFT frameworks [eq~\eqref{eq:ecsd2ineq}], while the strong-interaction ingredient, $W_\infty^\text{SD}$, has a different functional expression in each [eqs~\eqref{eq:WinfMP} and~\eqref{eq:Winftyksexp}], we choose to assess the performance of the LB AC integrand, $  W_\lam^\text{LB}$, while equating the two densities in the two frameworks, i.e. $\Delta n \equiv \Delta n^\hf$. In this way, the initial slope of the MP and DFT curve is forced to be the same. We do this to remove one dimension with respect to which the two exact MP and DFT AC integrands and/or their corresponding LB approximations may differ, such that their relevant features (their shape, how fast they go to their asymptotic value) be emphasized. This choice means that the external potential of the fully-interacting Hamiltonian is different in the two frameworks (because the HF density does not equal the interacting density for a given external potential).
In Figure~\ref{fig:WlambdaLB}, we plot the LB functional,  $ W_\lam^\text{LB}$, with input quantities $W_\infty^\hf$ or $W_\infty^\ks$, in contrast with the two exact AC integrands $W_\lam^\hf$ and $W_\lam^\ks$. This interpolation formula is clearly quite off the exact curve for the MP integrand in the weak-interaction regime ($\Delta n^\hf =1.9$), as the exact integrand changes curvature while the interpolant is convex [see panels (i) and (ii) of Figure~\ref{fig:WlambdaLB}]. The agreement between approximate and exact curves is qualitatively much better for intermediate and strong-interaction regimes ($\Delta n^\hf =1.0,\,0.1$).  The LB performance on the global correlation energy has already been investigated in ref~\onlinecite{GiaPri-JCP-22}. There, it was observed that the error $\Delta E_c^\text{SD} = E_c^\text{SD}-E_{c,\text{LB}}^\text{SD}$ is typically much smaller in the weak-interaction regime rather than in the strong-interaction and typically smaller in the MP framework rather than in the DFT (with the exception of a limited range around $U\approx\Delta v$, where $|\Delta E_c^\ks|<|\Delta E_c^\hf|$, compare Figure~11 of ref~\onlinecite{GiaPri-JCP-22}). However, inspection of panel (i) of Figure~\ref{fig:WlambdaLB} tells us that even in the $0<\lam<1$ range where the presence of the DCOC does not ``completely break'' the validity of the interpolant, it has limited ability to accurately model the exact integrand. Indeed, a quantity other than the LB error itself that is possibly more insightful is its relative error,
\begin{equation}\label{eq:relerrLB}
\text{rel. err. LB(SD)}=\frac{\Delta E_c^\text{SD}}{E_c^\text{SD}}.
\end{equation}
Plotting this quantity, as done in Figure~\ref{fig:relerrLB}, we immediately see the problematic aspects in using a convex interpolation formula to approximate the HF correlation energy: while the relative error in the DFT framework stays everywhere below 35\%, and its maximum decreases from $U=5$ to $U=10$, for the HF correlation energy the magnitude of the relative error appears to increase proportionally with $U$ in the region near $|\Delta v| \approx U$. The change in sign of the error is reflected in the fact that the approximate integrand lies below the exact one in panel (i) of Figure~\ref{fig:WlambdaLB} and above the exact ones in panel (iii) and (v). Interestingly, in the DFT framework, the interpolant appears to lie everywhere above the exact curve, explaining why the LB error on the KS correlation energy does not change sign.

%relerrLB

%\clearpage
\bibliographystyle{achemso}
\bibliography{MPDFTAC}
\end{document}